\documentclass[aip,jcp,reprint,nobalancelastpage]{revtex4-2}
\usepackage[T1]{fontenc}
\usepackage[utf8]{inputenc}
\usepackage{newtxtext}
\usepackage{mathtools}
\usepackage{newtxmath}
\usepackage{graphicx}
\usepackage{enumerate}

\DeclarePairedDelimiter{\set}{\{}{\}}
\DeclareMathOperator{\integral}{\mathbf{I}}
\DeclareMathOperator{\E}{\mathbf{E}}
\DeclareMathOperator{\prob}{\mathbf{P}}
\newcommand*{\odif}[1]{\,\mathrm{d} #1}
\newcommand*{\ind}[1]{\mathbf{1}_{#1}}
\newcommand*{\dt}{\Delta}
\newcommand*{\X}[1]{X_{#1}}
\newcommand*{\Y}[1]{Y_{#1}}
\newcommand*{\Z}[1]{Z_{#1}}
\newcommand*{\Xinf}{X}
\newcommand*{\Yinf}{Y}
\newcommand*{\Zinf}{Z}
\newcommand*{\Ix}{\integral^{X}}
\newcommand*{\Ixt}{\integral^{X,t}}
\newcommand*{\Iz}{\integral^{Z}}
\newcommand*{\Izt}{\integral^{Z,t}}
\newcommand*{\Iwx}{\integral_\rw^{X}}
\newcommand*{\Iwz}{\integral_\rw^{Z}}
\newcommand*{\Ex}{\E^{X}}
\newcommand*{\Ez}{\E^{Z}}

\newcommand*{\Ewx}{\E_\rw^{X}}
\newcommand*{\Ewz}{\E_\rw^{Z}}
\newcommand*{\Px}{\prob^{X}}
\newcommand*{\Pz}{\prob^{Z}}
\newcommand*{\rw}{\omega}
\newcommand*{\fn}{\eta}
\newcommand*{\pairfn}{\gamma}
\newcommand*{\stm}[1]{{\tau_-({#1})}}
\newcommand*{\stp}[1]{{\tau_+({#1})}}
\newcommand*{\qm}{q_-}
\newcommand*{\qp}{q_+}
\newcommand*{\py}[2]{\Omega(#1\,|\,#2)}
\newcommand*{\pyz}[1]{\Omega(\Y{#1}\,|\,\X{#1})}
\newcommand*{\pyzinf}{\Omega(\Yinf\,|\,\Xinf)}
\newcommand*{\xk}[2]{\kappa(#1\,|\,#2)}
\newcommand*{\xkz}[1]{\kappa(\Y{#1}\,|\,\X{#1})}
\newcommand*{\pymz}[1]{k_-(\Y{#1}\,|\,\X{#1})}
\newcommand*{\pypz}[1]{k_+(\Y{#1}\,|\,\X{#1})}
\newcommand*{\rate}{R}
\newcommand*{\density}[1]{\rho_{#1}}
\newcommand*{\current}[1]{J_{#1}}
\newcommand*{\projw}[1]{A_{#1}}
\newcommand*{\mtm}{m_-}
\newcommand*{\mtp}{m_+}
\newcommand*{\cv}{\theta}
\newcommand*{\cva}{{\cv'}}

\newcommand*{\cvpt}{v}
\newcommand*{\cvpta}{{\cvpt'}}
\newcommand*{\coca}{\zeta}

\begin{document}
\title{Augmented Transition Path Theory for Sequences of Events}
\author{Chatipat Lorpaiboon}
\affiliation{Department of Chemistry and James Franck Institute, University of Chicago, Chicago IL 60637, USA}
\author{Jonathan Weare}
\affiliation{Courant Institute of Mathematical Sciences, New York University, New York 10012, USA}
\author{Aaron R. Dinner}
\email{dinner@uchicago.edu}
\affiliation{Department of Chemistry and James Franck Institute, University of Chicago, Chicago IL 60637, USA}
\begin{abstract}
Transition path theory provides a statistical description of the dynamics of a reaction in terms of local spatial quantities. In its original formulation, it is limited to reactions that consist of trajectories flowing from a reactant set $A$ to a product set $B$. We extend the basic concepts and principles of transition path theory to reactions in which trajectories exhibit a specified sequence of events and illustrate the utility of this generalization on examples.
\end{abstract}
\maketitle

\section{Introduction}

Many reactions studied today proceed through competing pathways. Understanding such reactions relies on being able to assess the relative importance of the competing pathways and how they contribute to overall rates. When the pathways are well separated, they can be treated independently, often by traditional theories that assume a well-defined activated complex (transition state) and a simple form for the underlying (free) energy landscape governing the dynamics \cite{hanggi1990,peters2017reaction}. However, when the (observed) dynamics are stochastic, the pathways of reactions often overlap in configuration space. Approaches that treat competing pathways in a unified fashion are thus needed.

To this end, here, we build on transition path theory (TPT) \cite{e2006,vanden-eijnden2006,e2010,metzner2009,metzner2006}. The core idea of transition path theory is that the statistics of the ensemble of reactive trajectories can be related to quantities that are local in space: probability currents of reactive trajectories (henceforth, reactive currents) and committors. These quantities enable TPT to go beyond traditional theories by providing information about mechanisms. Reactive currents quantify flows in phase space. Committors, which are probabilities of reaching one metastable state before another, by definition characterize the progress of stochastic reactions \cite{du1998transition}.

In its traditional formulation, TPT focuses on transitions between two metastable states. In the present paper, we extend TPT to compute statistics for sequences of events, and we show how this significantly expands its applicability. Our work builds on but goes beyond previous studies. It is closely related to history-augmented Markov state models, in which states are labeled based on the last metastable state visited \cite{suarez2016}. Separating the ensemble of reactive trajectories using these labels enables rates to be computed from the flux into a metastable state \cite{vanden2009exact,dickson2009separating}, as well as reactive currents and committors from the underlying trajectories \cite{vani2022}. Our approach generalizes the labeling strategy to sequences of arbitrary numbers of states and allows specification of not just past events but also future ones.

Our work also has connections to that of Koltai and co-workers, who extended TPT to allow trajectories to leave and enter the region connecting the metastable states to analyze trajectory segments from satellite data for drifters in the ocean \cite{miron2021}. Specifically, they redefined committors to exclude trajectories that were not wholly within this region and noted that this approach could be used to exclude trajectories that pass through selected states. They also considered computing statistics for trajectories beginning and ending in specific portions of metastable states. Both of these developments allow statistics to be computed for subsets of reactive trajectories based on the states that they visit.

We present our work as follows. First, in Sections \ref{sec:ensemble} and \ref{sec:tpt}, we review how TPT expresses path statistics in terms of spatially and temporally local quantities using committors. In Section~\ref{sec:motivating}, we present a motivating example in which this is not possible within the existing framework, but it can be made possible by augmenting the stochastic process with labels that account for sequences of events. This is the key idea of the paper.  
While this idea is straightforward, formulating the theory in full generality requires some technical development, and readers may wish to skim Sections~\ref{sec:atpt} to \ref{sec:k} initially, focusing on the brief summaries in the first paragraphs of Sections~\ref{sec:atpt} and \ref{sec:k}.
In Section~\ref{sec:atpt}, we discuss conditions of consistency and Markovianity that must be satisfied for TPT to apply to the augmented process.  We also generalize committors and integrals based on them.  In Section~\ref{sec:stat}, we review the most commonly computed TPT statistics and show how they can be computed in our augmented TPT framework. In Section~\ref{sec:k}, we introduce a procedure for constructing the augmented process from pairs of successive time points rather than full trajectory segments, and we show how processes can be composed to construct more complex ones.
We summarize the operational procedure in Section \ref{sec:compute}.
Then, in Section~\ref{sec:example}, we illustrate our approach on two systems with multiple pathways and intermediates. In Section~\ref{sec:discussion}, possible extensions and numerical strategies for treating more complicated systems are discussed.
In Appendix~\ref{appendix:fdm}, we provide a method for calculating augmented TPT statistics using a finite difference scheme. Code implementing this method is available at \texttt{github.com/dinner-group/atpt}.

\section{Framework}

In this section, we review TPT to show how it casts statistics for reactive trajectories in terms of local quantities. Then we present an example that cannot be treated within the traditional TPT framework and show how it can be treated by introducing an augmented process. The essential idea is that the augmented process accounts for the order of events. The challenge in implementing this idea is that, for a finite-length trajectory segment, we generally do not know the events that occur before and after it.

For clarity, we present our results in terms of a discrete-time Markov process $\X{t}$ with time step $\dt$, but our results generalize to continuous-time processes in the limit $\dt \rightarrow 0$. We denote the time interval $r,r+\dt,\ldots,s$ by $r:s$ and a trajectory segment on this time interval by $\X{r:s} = (\X{r},\X{r+\dt},\ldots,\X{s})$. For conciseness, we denote an infinite trajectory $\X{-\infty:\infty}$ by $\Xinf$.

\subsection{Ensemble of Reactive Trajectories} \label{sec:ensemble}

In both traditional and augmented TPT, statistics are computed over the ensemble of reactive trajectories.  In this section, we define this ensemble and integrals over it.  Here, we focus on traditional TPT, but the framework generalizes to augmented TPT immediately once we define the augmented process in Section~\ref{sec:atpt}.

Traditional TPT considers a reaction from a set $A$ to a set $B$ via trajectories that cross a region $D$. In anticipation of our augmented framework, we allow $D \subseteq (A \cup B)^c$ as in ref.~\onlinecite{miron2021}. We consider a trajectory $\X{r:s}$ to be reactive if its first time point $\X{r}$ is in the reactant set $A$, its last time point $\X{s}$ is in the product set $B$, and all intervening time points $\X{r+\dt:s-\dt}$ are in the region $D$. Mathematically, we implement this definition through the indicator function
\begin{equation} \label{eq:tpt_w}
\rw(\X{r:s}) = \ind{A \times D \times \cdots \times D \times B}(\X{r:s}),
\end{equation}
where
\begin{equation}
\ind{S}(x) =
\begin{cases}
1 & \text{if } x \in S, \\
0 & \text{otherwise},
\end{cases}
\end{equation}
and $S_1 \times \cdots \times S_n = \set{(x_1,\ldots,x_n) \mid x_1 \in S_1, \ldots, x_n \in S_n}$ is the $n$-fold Cartesian product.

Given \eqref{eq:tpt_w}, we define the integral over the ensemble of reactive trajectories to be
\begin{equation} \label{eq:tpt_int}
\Iwx[\fn] = \lim_{T \rightarrow \infty} \frac{1}{2T} \Ix\Bigl[\sum_{\substack{r=-T:T-\dt \\ s=r+\dt:T}} \rw(\X{r:s}) \fn(\X{r:s})\Bigr],
\end{equation}
where $\Ix[f(\Xinf)]$ is the integral of $f(\Xinf)$ over the distribution of infinite trajectories $\Xinf$, which we denote using the superscript $\Xinf$. When $\X{t}$ is a stationary ergodic process and $\Ix$ is the expectation $\Ex$ over the distribution of infinite trajectories, as in traditional TPT, we can compute $\Iwx[\fn]$ from a single infinite trajectory and so $\Ix$ can be omitted; however, this is not necessarily true for time-dependent processes, as in ref.~\onlinecite{helfmann2020}, or for augmented processes, as in this work. As $\X{t}$ is a Markov process, we can compute this integral by sampling configurations $\X{-T}$ from the distribution of states at time $-T$ and propagating until time $T$. The prefactor $1/(2T)$ ensures that $\Iwx[\fn]$ gives consistent results across different trajectory lengths $2T$. We can then calculate expectations over the ensemble of reactive trajectories as
\begin{equation}
\Ewx[\fn] = \Iwx[\fn]/\Iwx[1],
\end{equation}
where the normalization factor $\Iwx[1]$ is the expected number of reactive trajectories which start (or end) per unit time. The integral $\Iwx[\fn]$ thus yields statistics that can be used to characterize and compare reaction pathways.

\subsection{Transition Path Theory} \label{sec:tpt}

In general, the ensemble of reactive trajectories can only be meaningfully interpreted through its statistics. Although these statistics can be computed directly from the ensemble of reactive trajectories, TPT enables them to be computed from other data as well by expressing them in terms of spatially and temporally local quantities.  

TPT specifically considers functions that can be written as
\begin{equation} \label{eq:functopair}
\fn(\X{r:s}) = \sum_{t=r:s-\dt} \pairfn(\X{t:t+\dt}) \dt,
\end{equation}
where $\pairfn(\X{t:t+\dt})$ is a function of successive time points $\X{t}$ and $\X{t+\dt}$. In this case, substituting \eqref{eq:functopair} into \eqref{eq:tpt_int} and exchanging the order of the sums yields
\begin{align}
\Iwx[\fn]
& = \lim_{T \rightarrow \infty} \frac{\dt}{2T} \sum_{t=-T:T-\dt} \Ix\Bigl[\sum_{\substack{r=-T:t \\ s=t+\dt:T}} \rw(\X{r:s}) \pairfn(\X{t:t+\dt})\Bigr]
\label{eq:tpt_substitute} \\
& = \Ixt\Bigl[\sum_{\substack{r=-\infty:t \\ s=t+\dt:\infty}} \rw(\X{r:s}) \pairfn(\X{t:t+\dt})\Bigr],
\label{eq:tpt_intshift}
\end{align}
where from \eqref{eq:tpt_substitute} to \eqref{eq:tpt_intshift} we have taken the limit $T \rightarrow \infty$ and performed a time average over $t$, which we denote by the superscript $t$. That is,
\begin{equation}
\Ixt[f(\Xinf,t)] = \lim_{T \rightarrow \infty} \frac{\dt}{2T} \sum_{t=-T:T-\dt} \Ix[f(\Xinf,t)].
\end{equation}

We can then factor
\begin{equation} \label{eq:factor}
\sum_{\substack{r=-\infty:t \\ s=t+\dt:\infty}} \rw(\X{r:s}) = \ind{A}(\X{\stm{t}}) \ind{B}(\X{\stp{t+\dt}}),
\end{equation}
where
\begin{align}
\stm{t} & = \max\set{t' \le t \mid \X{t'} \in D^c} \\
\stp{t} & = \min\set{t' \ge t \mid \X{t'} \in D^c}
\end{align}
are the last exit time from $D^c$ and the first entrance time to $D^c$, respectively. Equation \eqref{eq:factor} results from the identities
\begin{align}
\sum_{r=-\infty:t} \ind{A \times D \times \cdots \times D}(\X{r:t}) & = \ind{A}(\X{\stm{t}}),
\label{eq:tm_sum} \\
\sum_{s=t:\infty} \ind{D \times \cdots \times D \times B}(\X{t:s}) & = \ind{B}(\X{\stp{t}}).
\label{eq:tp_sum}
\end{align}
We arrive at \eqref{eq:tp_sum} by observing that only one term in the sum can be nonzero: because $D$ and $B$ are disjoint, $\ind{D \times \cdots \times D \times B}(\X{t:s})$ can be nonzero only when $s = \stp{t}$ is the first time $t' \ge t$ that $\X{t'} \notin D$. Similar logic applies for \eqref{eq:tm_sum}.

Consequently, for a Markov process, \eqref{eq:tpt_intshift} can be expressed in terms of only local quantities:
\begin{align}
\Iwx[\fn]
& = \Ixt[\ind{A}(\X{\stm{t}}) \ind{B}(\X{\stp{t+\dt}}) \pairfn(\X{t:t+\dt})]
\label{eq:tpt_split} \\
& = \Ixt[\qm(\X{t}) \qp(\X{t+\dt}) \pairfn(\X{t:t+\dt})],
\label{eq:tpt_exp}
\end{align}
where we have defined the backward and forward committors respectively as
\begin{align}
\qm(\X{t}) & = \Ex[\ind{A}(\X{\stm{t}}) \mid \X{t}],
\label{eq:qm} \\
\qp(\X{t}) & = \Ex[\ind{B}(\X{\stp{t}}) \mid \X{t}].
\label{eq:qp}
\end{align}
The backward committor $\qm(\X{t})$ is the probability that $\X{t}$ last came from $A$ rather than $(A \cup D)^c$, and the forward committor $\qp(\X{t})$ is the probability that $\X{t}$ will go to $B$ before $(B \cup D)^c$.

The main result of this section is \eqref{eq:tpt_exp}. The advantage of \eqref{eq:tpt_exp} over \eqref{eq:tpt_int} is that the former involves only statistics that are local in space and time. This aids in the interpretation of these statistics, and it enables their estimation from short trajectories $\X{t:t+\dt}$, thus eliminating the need for trajectories that actually cross from $A$ to $B$.

\subsection{A Motivating Reaction} \label{sec:motivating}

\begin{figure}
    \includegraphics{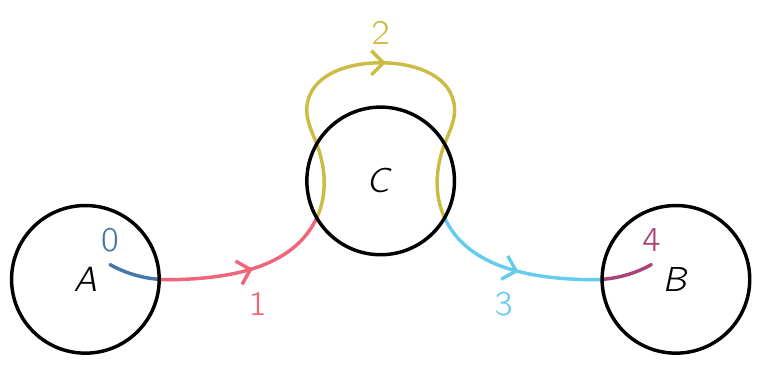}
    \caption{Values of $\Y{t}$ at each point of a reactive trajectory described by \eqref{eq:acb_w}.}
    \label{fig:traj}
\end{figure}

To motivate our augmented framework, we consider a reaction with an intermediate state $C$ and compute statistics only for reactive trajectories that proceed through the intermediate. The function that selects trajectories of interest is
\begin{equation} \label{eq:acb_w}
\rw(\X{r:s}) = \sum_{\substack{t_1=r+\dt:s-\dt \\ t_2=t_1:s-\dt}}
    \begin{aligned}[t]
    [ & \ind{A \times D \times \cdots \times D}(\X{r:t_1-\dt}) \\
    & \ind{C \times (C \cup D) \times \cdots \times (C \cup D) \times C}(\X{t_1:t_2}) \\
    & \ind{D \times \cdots \times D \times B}(\X{t_2+\dt:s})],
    \end{aligned}
\end{equation}
where $D = (A \cup B \cup C)^c$, and the sum allows for any $t_1$ and $t_2$ satisfying $r < t_1 \le t_2 < s$. The sum searches for times $t_1$ and $t_2$, which are the first and last times that the trajectory is in $C$. Because determining $t_1$ and $t_2$ requires a search over the entire trajectory $\X{r:s}$, we cannot factor $\rw(\X{r:s})$ as in \eqref{eq:factor}.

Now suppose that, for each reactive trajectory segment $\X{r:s}$ in the infinite trajectory $\Xinf$, we have a process $\Y{t}$ with
\begin{equation} \label{eq:acb_y}
\Y{t} =
    \begin{cases}
    0 & \text{if } t \le r, \\
    1 & \text{if } r < t < t_1, \\
    2 & \text{if } t_1 \le t \le t_2, \\
    3 & \text{if } t_2 < t < s, \\
    4 & \text{if } s \le t.
    \end{cases}
\end{equation}
An example reactive trajectory labeled with $\Y{t}$ is shown in Figure~\ref{fig:traj}. We can apply TPT on the augmented process $\Z{t} = (\X{t},\Y{t})$ because we can write \eqref{eq:acb_w} in the form of \eqref{eq:tpt_w} as
\begin{equation}
\rw(\Z{r:s}) = \ind{(A \times \set{0}) \times D' \times \cdots \times D' \times (B \times \set{4})}(\Z{r:s}),
\end{equation}
where $D' = (D \times \set{1}) \cup ((C \cup D) \times \set{2}) \cup (D \times \set{3})$.

This approach suggests a general strategy. We identify events---in this case, the first time $t_1$ and last time $t_2$ that $\X{t}$ is in $C$---and define a process $\Y{t}$ which labels these events. Then, we define reactive trajectories on the augmented state space using \eqref{eq:tpt_w}. So long as $\Z{t}$ satisfies the assumptions behind TPT, we can express statistics using local quantities in the same manner as in \eqref{eq:tpt_exp}.
In Section \ref{sec:atpt}, we discuss conditions for this to be the case, allowing for the possibility that an infinite trajectory has multiple labelings (e.g., to account for multiple finite reactive segments).

\subsection{Augmented Transition Path Theory} \label{sec:atpt}

In this section, we introduce a function $\pyzinf$ for constructing an ensemble of trajectories augmented with labels from the distribution of trajectories $\Xinf$. 
We first consider the case of infinite length trajectories $\Zinf=(\Xinf,\Yinf)$ and then the case of finite length trajectories $\Z{r:s}=(\X{r:s},\Y{r:s})$, to which one is limited in practice.
We discuss two conditions that must hold for our framework.  First, $\pyz{r:s}$ must be consistent with $\pyzinf$.
Second, $\Z{t}$ must be Markovian.
Later, in Section~\ref{sec:k}, we detail a specific construction of $\Yinf$ from $\Xinf$ which requires examining only successive pairs of time points $\X{t:t+\dt}$ and $\Y{t:t+\dt}$.

We now present our augmented framework. We replace $\X{t}$ with the augmented process $\Z{t} = (\X{t},\Y{t})$, where $\Y{t}$ augments $\X{t}$ with information about past and future events. Often, a single $\Yinf$ is associated with each infinite trajectory $\Xinf$ because the latter contains full information about the past and future of any $\X{t}$. However, cases arise in which multiple $\Yinf$ can be associated with a given infinite trajectory $\Xinf$. For example, in the motivating reaction above, we define a $\Yinf$ for each reactive trajectory segment in $\Xinf$ (i.e., we consider multiple $r$ and $s$). It is thus necessary to consider a distribution of $\Yinf$, and we compute integrals over the distribution of infinite trajectories $\Zinf$ as
\begin{equation}
\Iz[f(\Zinf)] = \Ix\Bigl[\int \pyzinf f(\Zinf) \odif{\Yinf}\Bigr],
\end{equation}
where $\pyzinf$ is the distribution of $\Yinf$ (and thus $\Zinf$) for a given infinite trajectory $\Xinf$.

This immediately yields analogues of \eqref{eq:tpt_exp}, \eqref{eq:qm}, and \eqref{eq:qp}:
\begin{align}
\Iwz[\fn] & = \Izt[\qm(\Z{t}) \qp(\Z{t+\dt}) \pairfn(\Z{t:t+\dt})],
\label{eq:atpt_tpt_exp} \\
\qm(\Z{t}) & = \Ez[\ind{A}(\Z{\stm{t}}) \mid \Z{t}],
\label{eq:atpt_tpt_qm} \\
\qp(\Z{t}) & = \Ez[\ind{B}(\Z{\stp{t}}) \mid \Z{t}],
\label{eq:atpt_tpt_qp}
\end{align}
where sets $A$, $B$, and $D$ are now defined on the augmented state space, and $\stm{t}$ and $\stp{t}$ are now on the augmented process.

However, we cannot yet evaluate these integrals and expectations because $\Y{t}$ and thus each time point $\Z{t}$ depends on the infinite trajectory $\Xinf$. Instead, we must convert integrals over $\Zinf$ to integrals over $\Xinf$ using
\begin{equation} \label{eq:aug_exp}
\Iz[f(\Z{r:s})] = \Ix\Bigl[\int \pyz{r:s} f(\Z{r:s}) \odif{\Y{r:s}}\Bigr],
\end{equation}
where the weight of $\Y{r:s}$ (and thus $\Z{r:s}$) given the trajectory segment $\X{r:s}$ is
\begin{equation} \label{eq:p_def}
\pyz{r:s}
= \Ex\Bigl[\iint \pyzinf \odif{\Y{-\infty:r-\dt}} \odif{\Y{s+\dt:\infty}} \Bigm| \X{r:s}\Bigr].
\end{equation}
When a single $\Yinf$ is associated with each infinite trajectory $\Xinf$, $\pyz{r:s}$ is the probability of $\Y{r:s}$ given $\X{r:s}$ and so $\int \pyz{r:s} \odif{\Y{r:s}} = 1$; this is not true in the general case. Equation \eqref{eq:p_def} is the first requirement of our augmented framework: we must be able to convert expectations involving $\pyzinf$, which depend on the infinite trajectory $\Xinf$, to those involving $\pyz{r:s}$, which depend only on the finite trajectory $\X{r:s}$.

Using \eqref{eq:aug_exp}, we can then write \eqref{eq:atpt_tpt_exp}, \eqref{eq:atpt_tpt_qm}, and \eqref{eq:atpt_tpt_qp} as
\begin{align}
\Iwz[\fn] & =
    \begin{aligned}[t]
    \Ixt\Bigl[\int & \pyz{t:t+\dt} \qm(\Z{t}) \qp(\Z{t+\dt}) \\ & \pairfn(\Z{t:t+\dt}) \odif{\Y{t:t+\dt}}\Bigr],
    \end{aligned}
\label{eq:atpt_exp} \\
\qm(\Z{t})
& = \frac{\Ex[\int \pyz{-\infty:t} \ind{A}(\Z{\stm{t}}) \odif{\Y{-\infty:t-\dt}} \mid \X{t}]}{\Ex[\int \pyz{-\infty:t} \odif{\Y{-\infty:t-\dt}} \mid \X{t}]} \\
& = \frac{\Ex[\int \pyz{-\infty:t} \ind{A}(\Z{\stm{t}}) \odif{\Y{-\infty:t-\dt}} \mid \X{t}]}{\pyz{t}},
\label{eq:qm_defn} \\
\qp(\Z{t})
& = \frac{\Ex[\int \pyz{t:\infty} \ind{B}(\Z{\stp{t}}) \odif{\Y{t+\dt:\infty}} \mid \X{t}]}{\Ex[\int \pyz{t:\infty} \odif{\Y{t+\dt:\infty}} \mid \X{t}]} \\
& = \frac{\Ex[\int \pyz{t:\infty} \ind{B}(\Z{\stp{t}}) \odif{\Y{t+\dt:\infty}} \mid \X{t}]}{\pyz{t}}.
\label{eq:qp_defn}
\end{align}
For $\qm(\Z{t})$ and $\qp(\Z{t})$, we excluded $\Y{t}$ from the variables over which we integrate because we conditioned on it. We note that when $\rw(\Z{r:s})$ has no dependence on $\Y{r:s}$ [i.e., $\rw(\Z{r:s})$ = $\rw(\X{r:s})$] and there is a one-to-one correspondence between $\Xinf$ and $\Zinf$ [i.e., $\int \pyzinf \odif{\Yinf} = 1$], we can recover the traditional TPT committors as
\begin{align}
\qm(\X{t}) &= \int \pyz{t} \qm(\Z{t}) \odif{\Y{t}}, \\
\qp(\X{t}) &= \int \pyz{t} \qp(\Z{t}) \odif{\Y{t}}.
\end{align}

In traditional TPT, $\X{t}$ must be a Markov process so that, from \eqref{eq:tpt_split} to \eqref{eq:tpt_exp}, we could take expectations of $\ind{A}(\X{\stm{t}})$ and $\ind{B}(\X{\stp{t+\dt}})$ to obtain committors $\qm(\X{t})$ and $\qp(\X{t+\dt})$. For the augmented process $\Z{t}$ to be similarly treatable, we also require it to be a Markov process. This requirement may be surprising because $\Y{t}$ can depend on the future of $\X{t}$. This can be understood by observing that, for the augmented process, the probability distribution of $\X{t+\dt}$ depends on both $\X{t}$ and $\Y{t}$. For example, for $\qp(\Z{t})$ in \eqref{eq:atpt_tpt_qp}, the distribution of $\X{t+\dt:\infty}$ conditioned on $\Z{t}$ is not the same as that of $\X{t+\dt:\infty}$ conditioned on $\X{t}$ alone, since $\Y{t}$ specifies that $\X{t}$ must undergo certain events in the future.

Since $\X{t}$ and $\Z{t}$ are Markov processes, we can factor the path probabilities $\Px[\X{r:s}]$ and $\Pz[\Z{r:s}]$ of the original and augmented processes:
\begin{align}
\Px[\X{r:s}] & = \Px[\X{r}] \prod_{t=r:s-\dt} \Px[\X{t+\dt} \mid \X{t}] \\
& = \Px[\X{r}] \prod_{t=r:s-\dt} \frac{\Px[\X{t:t+\dt}]}{\Px[\X{t}]},
\label{eq:x_markov} \\
\Pz[\Z{r:s}] & = \Pz[\Z{r}] \prod_{t=r:s-\dt} \Pz[\Z{t+\dt} \mid \Z{t}] \\
& = \Pz[\Z{r}] \prod_{t=r:s-\dt} \frac{\Pz[\Z{t:t+\dt}]}{\Pz[\Z{t}]}.
\label{eq:xy_markov}
\end{align}
As above, the superscripts indicate distributions of infinite trajectories. Thus, for example,
\begin{equation}
\Px[\X{r:s}] = \iint \Px[\Xinf] \odif{\X{-\infty:r-\dt}} \odif{\X{s+\dt:\infty}}
\end{equation}
is the probability of observing $\X{r:s}$ with all possible semi-infinite segments $\X{-\infty:r-\dt}$ and $\X{s+\dt:\infty}$ before and after $r:s$, respectively; $\Px[\Xinf]$ is the probability of a specific infinite trajectory $\Xinf$. The probability distribution of $\Zinf$ is
\begin{equation}
\Pz[\Zinf] = \pyzinf \Px[\Xinf] / c
\end{equation}
with $c = \int \pyzinf \Px[\Xinf] \odif{\Zinf}$. Therefore, from \eqref{eq:p_def},
\begin{align}
\pyz{r:s}
& = \frac{\iint \pyzinf \Px[\Xinf] \odif{\Z{-\infty:r-\dt}} \odif{\Z{s+\dt:\infty}}}{\Px[\X{r:s}]} \\
& = \frac{\iint c \Pz[\Zinf] \odif{\Z{-\infty:r-\dt}} \odif{\Z{s+\dt:\infty}}}{\Px[\X{r:s}]} \\
& = c \frac{\Pz[\Z{r:s}]}{\Px[\X{r:s}]}.
\label{eq:pyfrac}
\end{align}
To compute $\pyz{r:s}$, we divide \eqref{eq:xy_markov} by \eqref{eq:x_markov} and then apply \eqref{eq:pyfrac}:
\begin{equation} \label{eq:p_markov}
\pyz{r:s} = \pyz{r} \prod_{t=r:s-\dt} \frac{\pyz{t:t+\dt}}{\pyz{t}}.
\end{equation}
This factorization is the second requirement of our augmented framework: we must be able to construct $\pyzinf$, which depends on the infinite trajectory $\Xinf$, from $\pyz{t:t+\dt}$, which can only depend on pairs of successive time points $\X{t:t+\dt}$.

\section{Reactive Statistics} \label{sec:stat}

In this section, we discuss TPT statistics that provide information about mechanisms. These include committors, the reactive flux, the reactive density, the reactive current, and expectations over reactive trajectories that they enable computing. We present expressions for augmented TPT in the form of \eqref{eq:atpt_tpt_exp}, which can be evaluated using \eqref{eq:atpt_exp}. The corresponding expressions for traditional TPT can be obtained by replacing $\Z{t}$ with $\X{t}$. The statistics are normalized so that different reactions that are specified through different $\rw(\Z{r:s})$ but calculated from the same distribution of infinite trajectories $\Xinf$ are directly comparable.

We note that augmented TPT is useful even for reactions which can be described using traditional TPT [i.e., $\rw(\Z{r:s}) = \rw(\X{r:s})$]. The augmented process allows reaction mechanisms to be resolved in more detail, since committors and other statistics can be calculated on points which depend on both past and future behaviors of trajectories. Furthermore, the addition of past and future information enables the calculation of statistics with $\fn(\X{r:s})$ no longer restricted to the form in \eqref{eq:functopair}.

Several of the statistics that we discuss yield quantities on points $\cvpt$ in a collective variable (CV) space $\cv$, which we indicate using the subscript $\cv$. We express these statistics on a CV space rather than the state space of $\Z{t}$ because, for complex systems, it is often the case that the full state space contains variables that are irrelevant to understanding the reaction. This is particularly true for the augmented state space, which must contain the information required to select reactive trajectories using \eqref{eq:tpt_w} and compute statistics using \eqref{eq:functopair}, both of which rely on $\Y{t}$ to obtain past or future information. Nevertheless, the theory holds for the choice $\cv(\Z{t}) = \Z{t}$.

\subsection{Reactive Flux}

The reactive flux $\rate = \Iwz[1]$ is the expected number of reactive trajectories which start (or end) per unit time. We can express the reactive flux in the form of \eqref{eq:atpt_tpt_exp} by choosing $\pairfn(\Z{t:t+\dt})$ so that $\fn(\Z{r:s}) = 1$ when $\Z{r:s}$ is reactive. Such choices of $\pairfn(\Z{t:t+\dt})$ include $\ind{A}(\Z{t})/\dt$ and $\ind{B}(\Z{t+\dt})/\dt$, which are nonzero only when $\Z{t:t+\dt}$ is the first or last step of the reactive trajectory, respectively. Consequently, we can compute the reactive flux using
\begin{align}
\rate
& = \Izt[\ind{A}(\Z{t}) \qp(\Z{t+\dt}) / \dt]
\label{eq:rate_a} \\
& = \Izt[\qm(\Z{t}) \ind{B}(\Z{t+\dt}) / \dt],
\label{eq:rate_b}
\end{align}
where we have applied the identities $\ind{A}(\Z{t}) \qm(\Z{t}) = \ind{A}(\Z{t})$ and $\ind{B}(\Z{t}) \qp(\Z{t}) = \ind{B}(\Z{t})$. Equation \eqref{eq:rate_a} counts the number of trajectories that exit $A$ in the time interval $\dt$ and then react; \eqref{eq:rate_b} is the analogue for trajectories entering $B$.

The reactive flux is of interest not only in its own right but also for calculating expectations over reactive trajectories:
\begin{equation}
\Ewz[\fn] = \Iwz[\fn] / \Iwz[1] = \Iwz[\fn] / \rate.
\end{equation}
For example, the duration $N(\Z{r:s}) = s-r$ of a trajectory can be expressed in the form of \eqref{eq:functopair} with $\pairfn(\Z{t:t+\dt}) = 1$, and so the expected duration of a reactive trajectory is
\begin{equation}
\Ewz[N] = \Izt[\qm(\Z{t}) \qp(\Z{t+\dt})] / \rate.
\end{equation}

\subsection{Reactive Density}

The reactive density is the distribution of configurations which belong to reactive trajectories. For a point $\cvpt$ in the CV space $\cv$, the reactive density $\density{\cv}(\cvpt)$ is the probability that $\cv(\Z{t}) = \cvpt$ and is part of a reactive trajectory. Equivalently, it is the expected fraction of time an infinite trajectory spends reactive at $\cvpt$. It can be expressed in the form of \eqref{eq:atpt_tpt_exp} as
\begin{equation}
\density{\cv}(\cvpt) = \Izt\biggl[\qm(\Z{t}) \qp(\Z{t+\dt}) \frac{\delta_\cvpt(\cv(\Z{t})) + \delta_\cvpt(\cv(\Z{t+\dt}))}{2}\biggr],
\end{equation}
where $\delta$ is the Dirac delta function. When computing the expectation, $\delta_\cvpt(\cv(\Z{t}))$ selects the points $\Z{t}$ with $\cv(\Z{t}) = \cvpt$. The term $(\delta_\cvpt(\cv(\Z{t}))+\delta_\cvpt(\cv(\Z{t+\dt})))/2$ corresponds to assuming that half of the time of each step $\Z{t:t+\dt}$ is spent in $\Z{t}$ and half of the time is spent in $\Z{t+\dt}$.

In turn, the reactive density can be used to evaluate \eqref{eq:atpt_tpt_exp} when $\pairfn(\Z{t:t+\dt}) = (f(\cv(\Z{t}))+f(\cv(\Z{t+\dt})))/2$ is a path-independent function on the CV space:
\begin{equation}
\Iwz[\fn] = \int \density{\cv}(\cvpt) f(\cvpt) \odif{\cvpt}.
\end{equation}
For instance, the expected fraction of time an infinite trajectory spends reactive can be obtained by setting $f(\cvpt) = 1$, so that
\begin{equation}
\Iwz[N] = \int \density{\cv}(\cvpt) \odif{\cvpt} = \Izt[\qm(\Z{t}) \qp(\Z{t+\dt})],
\end{equation}
where we have assumed the distribution of trajectories $X$ to be a probability distribution, so that $\Ix[1]=1$.

We note that when the CV space $\cv$ is contained in the CV space $\cva$, i.e., we can write $\cv(\Z{t}) = \coca(\cva(\Z{t}))$ for some $\coca(\cvpta)$, we can calculate $\density{\cv}(\cvpt)$ by projecting $\density{\cva}(\cvpta)$ onto $\cv$:
\begin{equation}
\density{\cv}(\cvpt) = \int \delta_\cvpt(\coca(\cvpta)) \density{\cva}(\cvpta) \odif{\cvpta}.
\end{equation}

We can do the same for functions $f(\cvpta)$ defined on the CV space $\cva$. We calculate $\projw{\cv}[f](\cvpt)$, the expected value of $f(\cvpta)$ at a point $\cvpt$ in the CV space $\cv$, conditioned on trajectories passing through that point being reactive, as
\begin{equation} \label{eq:condstat}
\projw{\cv}[f](\cvpt) = \frac{\int \delta_\cvpt(\coca(\cvpta)) \density{\cva}(\cvpta) f(\cvpta) \odif{\cvpta}}{\density{\cv}(\cvpt)}.
\end{equation}

We emphasize that $f(\cvpta)$ can use $\Y{t}$ to obtain information from the past and future of $\X{t}$, and so \eqref{eq:condstat} is significantly more powerful than its traditional TPT counterpart. For instance, we can calculate the conditional mean first passage time, the expected time it takes for $\Z{t}$ to hit $B$ given that $\Z{t}$ is part of a reactive trajectory, using \eqref{eq:condstat} as discussed further in Section~\ref{cmfpt}, whereas in traditional TPT we would need to employ a Feynman--Kac formula (e.g., see ref.~\onlinecite{finkel2021learning}).

\subsection{Reactive Current}

The reactive current $\current{\cv}(\cvpt)$ through a point $\cvpt$ in the CV space $\cv$ is the net flow of reactive trajectories within $\cv$ through $\cvpt$. It can be expressed in the form of \eqref{eq:atpt_tpt_exp} as
\begin{equation} \label{eq:current}
\current{\cv}(\cvpt) =
\begin{aligned}[t]
\Izt\biggl[
& \qm(\Z{t}) \qp(\Z{t+\dt}) \\
& \frac{\delta_\cvpt(\cv(\Z{t}))+\delta_\cvpt(\cv(\Z{t+\dt}))}{2} \frac{\cv(\Z{t+\dt})-\cv(\Z{t})}{\dt}
\biggr].
\end{aligned}
\end{equation}
Conceptually, for each pair of successive time points $\Z{t:t+\dt}$ that is part of a reactive trajectory, we compute the numerical derivative $(\cv(\Z{t+\dt})-\cv(\Z{t}))/\dt$ and then split it equally between $\Z{t}$ and $\Z{t+\dt}$. In fact, in the limit $\dt \rightarrow 0$, when $\cv(\Z{t})$ is differentiable, equation \eqref{eq:current} becomes
\begin{equation}
\current{\cv}(\cvpt) = \Izt\biggl[\ind{A}(\Z{\stm{t}}) \ind{B}(\Z{\stp{t}}) \delta_\cvpt(\cv(\Z{t})) \frac{\mathrm{d} \cv(\Z{t})}{\mathrm{d} t}\biggr],
\end{equation}
which is the time derivative of $\cv(\Z{t})$ integrated over the distribution of reactive trajectories $\Z{t}$ with $\cv(\Z{t}) = \cvpt$.

It can be useful to compute the reactive current along the gradient of a function $f(\cvpt)$,
\begin{equation} \label{eq:current_op}
\current{\cv}[f](\cvpt) =
\begin{aligned}[t]
\Izt\biggl[
& \qm(\Z{t}) \qp(\Z{t+\dt}) \\
& \frac{\delta_\cvpt(\cv(\Z{t}))+\delta_\cvpt(\cv(\Z{t+\dt}))}{2} \\
& \frac{f(\cv(\Z{t+\dt}))-f(\cv(\Z{t}))}{\dt}\biggr].
\end{aligned}
\end{equation}
In the limit $\dt \rightarrow 0$, for differentiable $f(\cvpt)$, equation \eqref{eq:current_op} is $\current{\cv}[f](\cvpt) = \current{\cv}(\cvpt) \cdot \nabla_\cv f(\cvpt)$ which we can derive by observing that the finite differences in \eqref{eq:current} and \eqref{eq:current_op} are ${\mathrm{d} \cv(\Z{t})}/{\mathrm{d} t}$ and ${\mathrm{d} f(\cv(\Z{t}))}/{\mathrm{d} t}$ in this limit, respectively, and by the chain rule ${\mathrm{d} f(\cv(\Z{t}))}/{\mathrm{d} t} = {\mathrm{d} \cv(\Z{t})}/{\mathrm{d} t} \cdot \nabla_\cv f(\cv(\Z{t}))$.

Like the reactive density, we can calculate $\current{\cv}(\cvpt)$ and $\current{\cv}[f](\cvpt)$ by projecting $\current{\cva}(\cvpta)$ and $\current{\cva}[f \circ \coca](\cvpta)$ onto $\cv$:
\begin{align}
\current{\cv}(\cvpt) & = \int \delta_\cvpt(\coca(\cvpta)) \current{\cva}[\coca](\cvpta) \odif{\cvpta}, \\
\current{\cv}[f](\cvpt) & = \int \delta_\cvpt(\coca(\cvpta)) \current{\cva}[f \circ \coca](\cvpta) \odif{\cvpta},
\end{align}
where $(f \circ \coca)(\cvpta) = f(\coca(\cvpta))$.

\subsection{Committors}

The committors $\qm(\Z{t})$ and $\qp(\Z{t})$ are defined on the state space of $\Z{t}$, which makes them useful for calculating other statistics but can make them hard to interpret. To address this issue, we can treat the committors as reaction coordinates, and project them onto a CV space $\cv$ as $\projw{\cv}[\qm](\cvpt)$ and $\projw{\cv}[\qp](\cvpt)$. These quantities have a physical interpretation. For instance, $\projw{\cv}[\qp](\cvpt)$ is the probability that a trajectory starting at a point $\Z{t}$ that is drawn from configurations with $\cv(\Z{t}) = \cvpt$ in the ensemble of reactive trajectories, will enter $B$ when it first leaves $D$. We note that, unlike most other reactive statistics, $\projw{\cv}[\qm](\cvpt)$ and $\projw{\cv}[\qp](\cvpt)$ with $\cv(\Z{t}) = \X{t}$ are not independent of the choice of $\Y{t}$, even when the same ensemble of reactive trajectories is selected because the likelihood that a trajectory contributes positively to the committor and the likelihood that it is reactive are correlated.

\subsection{Conditional Mean First and Last Passage Times} \label{cmfpt}

The first passage time to the product is the time it takes for a trajectory starting at time $t$ to reach the product $B$, at time $\stp{t}$. It can be expressed as
\begin{equation} \label{eq:ytp}
f(\Z{t}') = \Y{t}' =
\begin{cases}
\Y{t+\dt}' + \dt & \text{if } \Z{t} \notin B, \\
0 & \text{otherwise},
\end{cases}
\end{equation}
where $\Z{t}' = (\Z{t},\Y{t}')$. This increments $\Y{t}'$ by $\dt$ for each time step backward in time when $\Z{t} \notin B$, and sets $\Y{t}' = 0$ when $\Z{t} \in B$. The conditional mean first passage time, $\mtp(\Z{t})$, is the expected first passage time to the product for a point $\Z{t}$ that is part of a reactive trajectory. This statistic, and higher moments of the first passage time distribution, are useful for real-time forecasting, e.g., of weather \cite{finkel2021learning}. To compute the conditional mean first passage time, we take the conditional expectation of \eqref{eq:ytp} with respect to the reactive density using \eqref{eq:condstat}, i.e.,
\begin{equation}
\mtp(\Z{t}) = \projw{\cv'}[f](\Z{t}),
\end{equation}
where $\cv'(\Z{t}') = \Z{t}$. We can also calculate more general statistics on the distribution of the first passage time. For instance, the conditional variance of the first passage time to the product is $\projw{\cv'}[f^2](\Z{t}) - (\projw{\cv'}[f](\Z{t}))^2$. Likewise, the last passage time from the reactant is the time it takes for a trajectory ending at time $t$ to come from the reactant $A$, at time $\stm{t}$, conditioned on $\Z{t}$ being part of a reactive trajectory, and can be expressed as
\begin{equation} \label{eq:ytm}
g(\Z{t}'') = \Y{t}'' =
\begin{cases}
\Y{t-\dt}'' + \dt & \text{if } \Z{t} \notin A, \\
0 & \text{otherwise},
\end{cases}
\end{equation}
where $\Z{t}'' = (\Z{t},\Y{t}'')$. The conditional mean last passage time from the reactant is then
\begin{equation}
\mtm(\Z{t}) = \projw{\cv''}[g](\Z{t}),
\end{equation}
where $\cv''(\Z{t}'') = \Z{t}$. These statistics can be projected onto points $\cvpt$ on a CV space $\cv(\Z{t})$ as $\projw{\cv}[\mtm](\cvpt)$ and $\projw{\cv}[\mtp](\cvpt)$.

\section{Construction of the Augmented Process} \label{sec:k}

In this section, we describe a particularly useful way to define the augmented process. Namely, we decompose $\py{\Yinf}{\Xinf}$ into a product over functions of successive time points,  $\xkz{t:t+\dt}$. We show that $\py{\Y{r:s}}{\X{r:s}}$ so defined satisfies the required properties \eqref{eq:p_def} and \eqref{eq:p_markov} by construction.  We then demonstrate the construction of complex augmented processes from simpler ones by composition.  Finally, we show how this machinery applies to the motivating reaction.

\subsection{Decomposition of $\Omega$}

\begin{figure*}
    \includegraphics{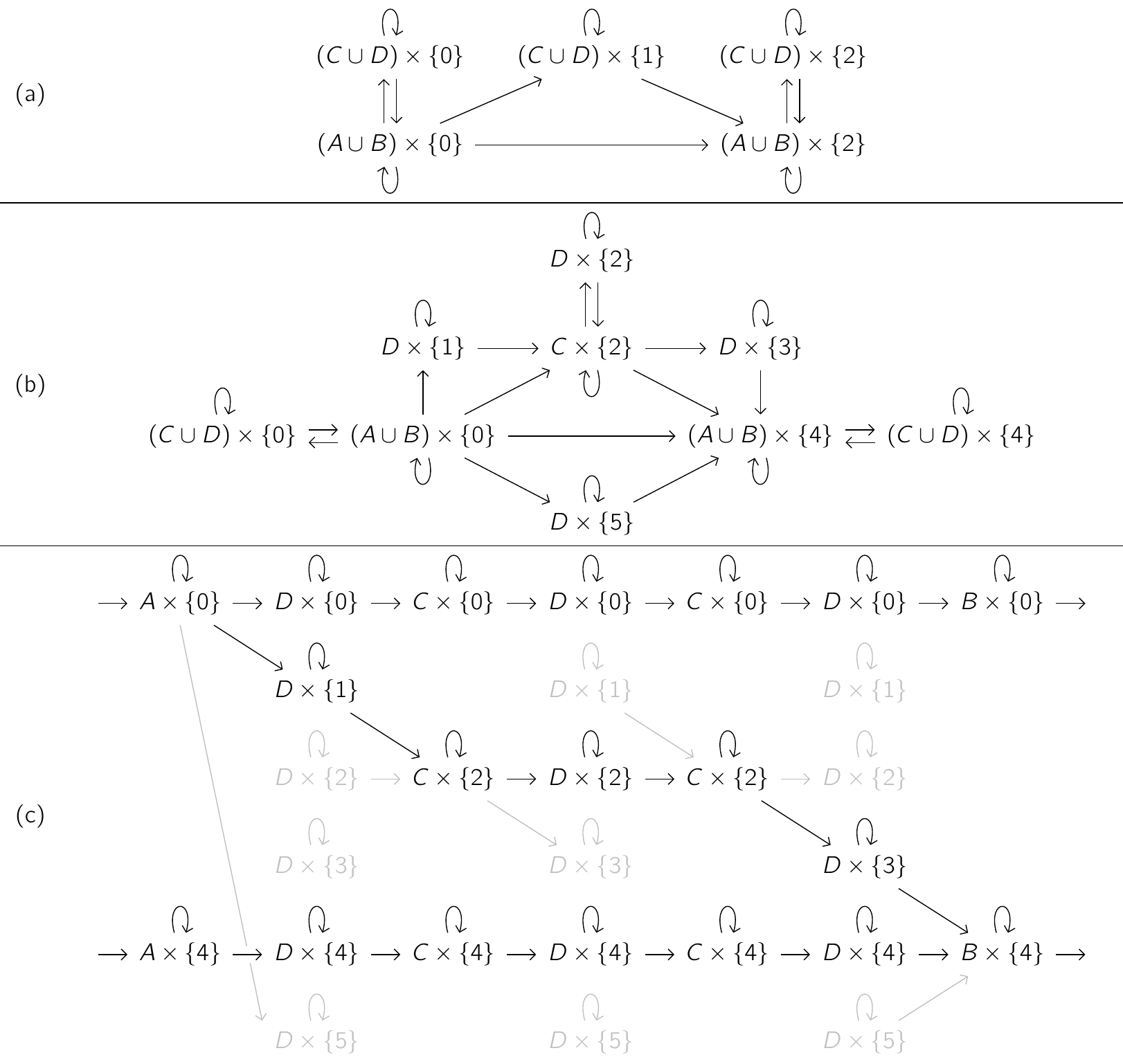}
    \caption{
        Construction of the augmented process for a reaction with an intermediate.
        (a) Possible transitions of the augmented process defined by \eqref{eq:y1}.
        (b) Possible transitions of the augmented process defined by \eqref{eq:k}. Each arrow from one set to another indicates that $\Y{t}$ may transition to $\Y{t+\dt}$ when $\Z{t}$ is in the first set and $\Z{t+\dt}$ is in the second set.
        (c) Determination of $\Y{r:s}$ for the reactive trajectory $\X{r:s}$ from Figure~\ref{fig:traj}. The black elements indicate trajectories $\Y{r:s}$ that satisfy $\xkz{r:s} = 1$. The gray elements indicate pairs $\Y{t:t+\dt}$ that satisfy $\xkz{t:t+\dt}=1$ but do not belong to any $\Y{r:s}$ with $\xkz{r:s}$.
    }
    \label{fig:y}
\end{figure*}

To apply our augmented framework, we need to construct the augmented process and in turn the distribution of infinite trajectories $\Zinf$ such that $\pyz{r:s}$ satisfies \eqref{eq:p_def} and \eqref{eq:p_markov}. One way to do so is to define $\pyzinf$, calculate $\pyz{r:s}$ from $\pyzinf$ using \eqref{eq:p_def}, and then verify that \eqref{eq:p_markov} holds. In this section, we present an alternative approach. We specify the augmented process through a function of pairs of successive time points, $\xkz{t:t+\dt}$, and then use it to calculate $\pyz{r:s}$. This procedure satisfies \eqref{eq:p_def} and \eqref{eq:p_markov} by construction.

We start by using the pair structure of \eqref{eq:p_markov} to factor
\begin{align} \label{eq:p_pairs}
\pyzinf = \prod_{t=-\infty:\infty} \xkz{t:t+\dt}.
\end{align}
It follows immediately that
\begin{equation} \label{eq:k_factor}
\pyzinf = \xkz{-\infty:r} \xkz{r:s} \xkz{s:\infty},
\end{equation}
where
\begin{equation}
\xkz{r:s} = \prod_{t=r:s-\dt} \xkz{t:t+\dt}.
\end{equation}
This cleanly separates terms which depend on the past $\X{-\infty:r}$ and future $\X{s:\infty}$ from those which depend on the trajectory segment $\X{r:s}$.

Using \eqref{eq:k_factor}, we can compute \eqref{eq:p_def} as
\begin{equation} \label{eq:p_traj}
\pyz{r:s} = \pymz{r} \xkz{r:s} \pypz{s},
\end{equation}
where we have defined
\begin{align}
\pymz{t} & = \Ex\Bigl[
    \int \xkz{-\infty:t} \odif{\Y{-\infty:t-\dt}}
    \Bigm| \X{t}\Bigr],
\label{eq:pm} \\
\pypz{t} & = \Ex\Bigl[
    \int \xkz{t:\infty} \odif{\Y{t+\dt:\infty}}
    \Bigm| \X{t}\Bigr].
\label{eq:pp}
\end{align}
We describe how to calculate $\pymz{t}$ and $\pypz{t}$ in Section~\ref{sec:compute}. The case $r = s = t$ is the weight of $\Y{t}$ given $\X{t}$:
\begin{equation} \label{eq:p_frame}
\pyz{t} = \pymz{t} \pypz{t}.
\end{equation}
We can verify that the resulting $\Z{t}$ is a Markov process by substituting \eqref{eq:p_traj} and \eqref{eq:p_frame} into \eqref{eq:p_markov}.

\subsection{Building Augmented Processes by Composition}

The factorization in \eqref{eq:p_pairs} has a number of advantages over \eqref{eq:p_markov}. One is that it facilitates deriving useful expressions for treating multiple augmented processes. If we have $\pyzinf$ of the form
\begin{equation}
\pyzinf = \prod_n \py{\Yinf^{(n)}}{\Xinf^{(n)}},
\end{equation}
where $n$ labels different augmented processes, we can factor both sides by \eqref{eq:p_pairs} to obtain
\begin{equation} \label{eq:k_combine}
\xkz{t:t+\dt} = \prod_n \xk{\Y{t:t+\dt}^{(n)}}{\X{t:t+\dt}^{(n)}},
\end{equation}
which involves only successive time points. Each of the terms $\xk{\Y{t:t+\dt}^{(n)}}{\X{t:t+\dt}^{(n)}}$ defines a process $\Y{t}^{(n)}$ using information from the original process $\X{t}$ and other processes $\Y{t}^{(m)}$, which we denote as $\X{t}^{(n)}$. We can use this to combine multiple augmented processes which may be defined independently or hierarchically.

For example, consider the augmented process $\Z{t}^\rw = (\X{t},\Y{t}^\rw)$, where $\Y{t}^\rw$ is used to define pathways and is defined using $\xk{\Y{t:t+\dt}^\rw}{\X{t:t+\dt}}$. To compute statistics on the first and last passage times, we can augment this process with the augmented processes in \eqref{eq:ytp} and \eqref{eq:ytm} by defining
\begin{align}
\xk{\Y{t:t+\dt}'}{\Z{t:t+\dt}^\rw} & =
\begin{cases}
    \delta_{\Y{t}'}({\Y{t+\dt}'} + \dt) & \text{if } \Z{t}^\rw \notin B, \\
    \delta_{\Y{t}'}(0) & \text{otherwise},
\end{cases} \\
\xk{\Y{t:t+\dt}''}{\Z{t:t+\dt}^\rw} & =
\begin{cases}
    \delta_{\Y{t+\dt}''}({\Y{t}''} + \dt) & \text{if } \Z{t+\dt}^\rw \notin B, \\
    \delta_{\Y{t+\dt}''}(0) & \text{otherwise}.
\end{cases}
\end{align}
The combined process $\Y{t} = (\Y{t}^\rw,\Y{t}',\Y{t}'')$ is then specified by
\begin{equation}
\begin{aligned}
\MoveEqLeft[1] \xkz{t:t+\dt} \\
& = \xk{\Y{t:t+\dt}^\rw}{\X{t:t+\dt}} \xk{\Y{t:t+\dt}'}{\Z{t:t+\dt}^\rw} \xk{\Y{t:t+\dt}''}{\Z{t:t+\dt}^\rw}.
\end{aligned}
\end{equation}
This example furthermore shows how \eqref{eq:p_pairs} allows forward-in-time and backward-in-time augmented processes to be treated in a unified manner and combined, which is not straightforward with \eqref{eq:p_markov}.

\subsection{Augmented Process for the Motivating Reaction}

We can also use \eqref{eq:k_combine} to construct augmented processes by combining simpler augmented processes. Here, we detail a possible construction of the augmented process \eqref{eq:acb_y} as a composite of three augmented processes.  First, we define an augmented process $\Z{t}^{(0)} = (\X{t},\Y{t}^{(0)})$ that selects all reactive trajectories regardless of pathway:
\begin{equation} \label{eq:y1}
\xk{\Y{t:t+\dt}^{(0)}}{\X{t:t+\dt}} =
\begin{aligned}[t]
[ & \ind{\set{0} \times \set{0}}(\Y{t:t+\dt}^{(0)}) \\
& + \ind{((D \cup C) \times \set{1}) \times ((D \cup C) \times \set{1})}(\Z{t:t+\dt}^{(0)}) \\
& + \ind{\set{2} \times \set{2}}(\Y{t:t+\dt}^{(0)}) \\
& + \ind{((A \cup B) \times \set{0}) \times ((D \cup C) \times \set{1})}(\Z{t:t+\dt}^{(0)}) \\
& + \ind{((D \cup C) \times \set{1}) \times ((A \cup B) \times \set{2})}(\Z{t:t+\dt}^{(0)}) \\
& + \ind{((A \cup B) \times \set{0}) \times ((A \cup B) \times \set{2})}(\Z{t:t+\dt}^{(0)}) ].
\end{aligned}
\end{equation}
For a reactive trajectory $\X{r:s}$ from $A$ to $B$, $\Y{t}^{(0)}$ splits the infinite trajectory $\Xinf$ into three parts: time points $\X{-\infty:r}$ before the reaction ($\Y{t}^{(0)} = 0$), time points $\X{r+\dt:s-\dt}$ during the reaction ($\Y{t}^{(0)} = 1$), and time points $\X{s:\infty}$ after the reaction ($\Y{t}^{(0)} = 2$). We list the possible transitions of this augmented process in Figure~\ref{fig:y}(a). The nodes are sets in which $\Z{t}^{(0)}$ may belong; an arrow from one set to another indicates that $\Y{t}^{(0)}$ may transition to $\Y{t+\dt}^{(0)}$ when $\Z{t}^{(0)}$ is in the first set and $\Z{t+\dt}^{(0)}$ is in the second set. For instance, the fourth term in \eqref{eq:y1} corresponds to the arrow from $(A \cup B) \times \set{0}$ to $(D \cup C) \times \set{1}$.

Next, we employ additional augmented processes to find the first and last times $t_1$ and $t_2$ that the trajectory is in $C$. Using \eqref{eq:y1}, we define the processes
\begin{align}
\Y{t}^{(1)} & =
\begin{cases}
\Y{t-\dt}^{(1)} & \text{if } \X{t} \in D \text{ and } \Y{t}^{(0)} = 1, \\
1 & \text{if } \X{t} \in C \text{ and } \Y{t}^{(0)} = 1, \\
0 & \text{if } \Y{t}^{(0)} \in \set{0,2},
\end{cases}
\label{eq:y2}
\\
\Y{t}^{(2)} & =
\begin{cases}
\Y{t+\dt}^{(2)} & \text{if } \X{t} \in D \text{ and } \Y{t}^{(0)} = 1, \\
1 & \text{if } \X{t} \in C \text{ and } \Y{t}^{(0)} = 1, \\
0 & \text{if } \Y{t}^{(0)} \in \set{0,2}.
\end{cases}
\label{eq:y3}
\end{align}
During the reaction (i.e., $\Y{t}^{(0)} = 1$), $\Y{t}^{(1)} = 1$ for times $t \ge t_1$ and $\Y{t}^{(2)} = 1$ for times $t \le t_2$. We can write \eqref{eq:y2} and \eqref{eq:y3} as
\begin{align}
\xk{\Y{t:t+\dt}^{(1)}}{\Z{t:t+\dt}^{(0)}} & =
\begin{aligned}[t]
[ & \ind{D \times \set{1} \times \set{(0,0),(1,1)}}(\X{t+\dt},\Y{t+\dt}^{(0)},\Y{t:t+\dt}^{(1)}) \\
& + \ind{C \times \set{1} \times \set{1}}(\X{t+\dt},\Y{t+\dt}^{(0)},\Y{t+\dt}^{(1)}) \\
& + \ind{\set{0,2} \times \set{0}}(\Y{t+\dt}^{(0)},\Y{t+\dt}^{(1)}) ],
\end{aligned}
\label{eq:k2} \\
\xk{\Y{t:t+\dt}^{(2)}}{\Z{t:t+\dt}^{(0)}} & =
\begin{aligned}[t]
[ & \ind{D \times \set{1} \times \set{(0,0),(1,1)}}(\X{t},\Y{t}^{(0)},\Y{t:t+\dt}^{(2)}) \\
& + \ind{C \times \set{1} \times \set{1}}(\X{t},\Y{t}^{(0)},\Y{t}^{(2)}) \\
& + \ind{\set{0,2} \times \set{0}}(\Y{t}^{(0)},\Y{t}^{(2)}) ].
\end{aligned}
\label{eq:k3}
\end{align}
We can then combine \eqref{eq:y1}, \eqref{eq:k2}, and \eqref{eq:k3} using \eqref{eq:k_combine}:
\begin{equation} \label{eq:k}
\xkz{t:t+\dt} =
\begin{aligned}[t]
[ & \ind{\set{0} \times \set{0}}(\Y{t:t+\dt}) \\
& + \ind{(D \times \set{1}) \times (D \times \set{1})}(\Z{t:t+\dt}) \\
& + \ind{((C \cup D) \times \set{2}) \times ((C \cup D) \times \set{2})}(\Z{t:t+\dt}) \\
& + \ind{(D \times \set{3}) \times (D \times \set{3})}(\Z{t:t+\dt}) \\
& + \ind{\set{4} \times \set{4}}(\Y{t:t+\dt}) \\
& + \ind{(D \times \set{5}) \times (D \times \set{5})}(\Z{t:t+\dt}) \\
& + \ind{((A \cup B) \times \set{0}) \times (D \times \set{1})}(\Z{t:t+\dt}) \\
& + \ind{((A \cup B) \times \set{0}) \times (C \times \set{2})}(\Z{t:t+\dt}) \\
& + \ind{(D \times \set{1}) \times (C \times \set{2})}(\Z{t:t+\dt}) \\
& + \ind{(C \times \set{2}) \times (D \times \set{3})}(\Z{t:t+\dt}) \\
& + \ind{(C \times \set{2}) \times ((A \cup B) \times \set{4})}(\Z{t:t+\dt}) \\
& + \ind{(D \times \set{3}) \times ((A \cup B) \times \set{4})}(\Z{t:t+\dt}) \\
& + \ind{((A \cup B) \times \set{0}) \times (D \times \set{5})}(\Z{t:t+\dt}) \\
& + \ind{(D \times \set{5}) \times ((A \cup B) \times \set{4})}(\Z{t:t+\dt}) \\
& + \ind{((A \cup B) \times \set{0}) \times ((A \cup B) \times \set{4})}(\Z{t:t+\dt}) ],
\end{aligned}
\end{equation}
where $\Z{t} = (\X{t},\Y{t})$, and to match \eqref{eq:acb_y}, we have merged $\Y{t}^{(0)}$, $\Y{t}^{(1)}$, and $\Y{t}^{(2)}$ into
\begin{equation}
\Y{t} =
\begin{cases}
0 & \text{if } (\Y{t}^{(0)},\Y{t}^{(1)},\Y{t}^{(2)}) = (0,0,0), \\
1 & \text{if } (\Y{t}^{(0)},\Y{t}^{(1)},\Y{t}^{(2)}) = (1,0,1), \\
2 & \text{if } (\Y{t}^{(0)},\Y{t}^{(1)},\Y{t}^{(2)}) = (1,1,1), \\
3 & \text{if } (\Y{t}^{(0)},\Y{t}^{(1)},\Y{t}^{(2)}) = (1,1,0), \\
4 & \text{if } (\Y{t}^{(0)},\Y{t}^{(1)},\Y{t}^{(2)}) = (2,0,0), \\
5 & \text{if } (\Y{t}^{(0)},\Y{t}^{(1)},\Y{t}^{(2)}) = (1,0,0).
\end{cases}
\end{equation}
We list the possible transitions of this augmented process in Figure~\ref{fig:y}(b).

In Figure~\ref{fig:y}(c), we illustrate the determination of $\Y{r:s}$ from $\xkz{t:t+\dt}$ for the trajectory $\X{r:s}$ in Figure~\ref{fig:traj}. For each time point $\X{t}$, we list the possible values of $\Y{t}$ from \eqref{eq:k}. We then stitch together trajectories by connecting successive time-point pairs $\Z{t:t+\dt}$ and following the arrows in Figure~\ref{fig:y}(b). The black elements in Figure~\ref{fig:y}(c) indicate step pairs $\Y{t:t+\dt}$ in trajectories $\Y{r:s}$ that satisfy $\xkz{r:s} = 1$, and the gray elements indicate step pairs $\Y{t:t+\dt}$ that satisfy $\xkz{t:t+\dt} = 1$ but do not belong to any $\Y{r:s}$ that satisfies $\xkz{r:s} = 1$.

There is one $\Yinf$ associated with each reactive trajectory segment in $\Xinf$. The diagonal path in Figure~\ref{fig:y}(c) represents one such reactive trajectory segment and can be selected using augmented TPT. The horizontal paths are collections of augmented processes corresponding to times before each future reaction ($\Y{t}=0$) and after each past reaction ($\Y{t}=4$).

\section{Algorithm} \label{sec:compute}

In this section, we summarize the operational aspects of the method. 
For the numerical examples that we consider in the present paper, we evaluate integrals of the form in \eqref{eq:tpt_int} using a finite difference approximation, which we detail in Appendix~\ref{appendix:fdm}; more complex systems can be treated by extending the approach in refs.~\onlinecite{thiede2019,strahan2021}, which we leave for future work.
In terms of the finite difference approximation, the algorithm for evaluating these statistics is as follows:
\begin{enumerate}[\hspace{0pt}1.]
\item Define $\xkz{t:t+\dt}$, $\rw(\Z{r:s})$, and $\pairfn(\Z{t:t+\dt})$ for the statistic of interest.
\item{Compute $\pymz{t}$ and $\pypz{t}$, which account for weights associated with the past and future segments of trajectories. To this end, we express \eqref{eq:pm} and \eqref{eq:pp} as
\begin{align}
\pymz{t} & =
\begin{aligned}[t]
\Ex\Bigl[\int & \xkz{t-\dt:t} \\
& \pymz{t-\dt} \odif{\Y{t-\dt}} \Bigm| \X{t}\Bigr],
\end{aligned}
\label{eq:pym_bvp}
\\
\pypz{t} & =
\begin{aligned}[t]
\Ex\Bigl[\int & \xkz{t:t+\dt} \\
& \pypz{t+\dt} \odif{\Y{t+\dt}} \Bigm| \X{t}\Bigr],
\end{aligned}
\label{eq:pyp_bvp}
\end{align}
and solve these equations using \eqref{eq:fdm_m} and \eqref{eq:fdm_p}.
}
\item Compute 
$\pyz{t:t+\dt}$ and $\pyz{t}$ by \eqref{eq:p_traj} and \eqref{eq:p_frame}, respectively.  
\item{Compute $\qm(\Z{t})$ and $\qp(\Z{t})$.  To this end, we express the committors as solutions to boundary value problems.  For $\Z{t}\in D$, 
\begin{align}
\qm(\Z{t})
& =
\Ex[\qm(\Z{t-\dt}) \mid \Z{t}]
\label{eq:qm_bvp} \\
& =
\begin{aligned}[t]
\Ex\Bigl[\int & \pyz{t-\dt:t} \\ 
& \qm(\Z{t-\dt}) \odif{\Y{t-\dt}} \Bigm| \X{t}\Bigr]\Bigm/\pyz{t},
\end{aligned}
\label{eq:xqm_bvp} \\
\qp(\Z{t}) & =
\Ex[\qp(\Z{t+\dt}) \mid \Z{t}]
\label{eq:qp_bvp} \\
& =
\begin{aligned}[t]
\Ex\Bigl[\int & \pyz{t:t+\dt} \\
& \qp(\Z{t+\dt}) \odif{\Y{t+\dt}} \Bigm| \X{t}\Bigr]\Bigm/\pyz{t}.
\end{aligned}
\label{eq:xqp_bvp}
\end{align}
For $\Z{t}\notin D$, $\qm(\Z{t}) = \ind{A}(\Z{t})$ and $\qp(\Z{t})= \ind{B}(\Z{t})$.
Above, \eqref{eq:qm_bvp} and \eqref{eq:qp_bvp} result from applying the identities \eqref{eq:tm_sum} and \eqref{eq:tp_sum} to the definitions \eqref{eq:qm} and \eqref{eq:qp}, and \eqref{eq:xqm_bvp} and \eqref{eq:xqp_bvp} follow in turn from \eqref{eq:qm_defn} and \eqref{eq:qp_defn}.  We solve \eqref{eq:xqm_bvp} and \eqref{eq:xqp_bvp} using \eqref{eq:fdm_m} and \eqref{eq:fdm_p}.
}
\item Evaluate \eqref{eq:atpt_exp} using \eqref{eq:fdm_exp}.
\end{enumerate}

\section{Numerical Examples} \label{sec:example}

In this section, we demonstrate our augmented framework on simple examples that make the limitations of traditional TPT apparent. The examples we consider employ overdamped Langevin dynamics on a potential $U(x)$ and satisfy the Fokker--Planck equation
\begin{equation} \label{eq:overdamped}
\frac{\partial \Px[\X{t}]}{\partial t} = \nabla \cdot (\Px[\X{t}] \nabla U(\X{t})) + \nabla^2 \Px[\X{t}].
\end{equation}
We calculate all statistics using a quadrature scheme adapted from ref.~\onlinecite{thiede2019}, which we detail in Appendix~\ref{appendix:fdm}.

\subsection{Reaction through an Intermediate}

\begin{figure}
    \includegraphics{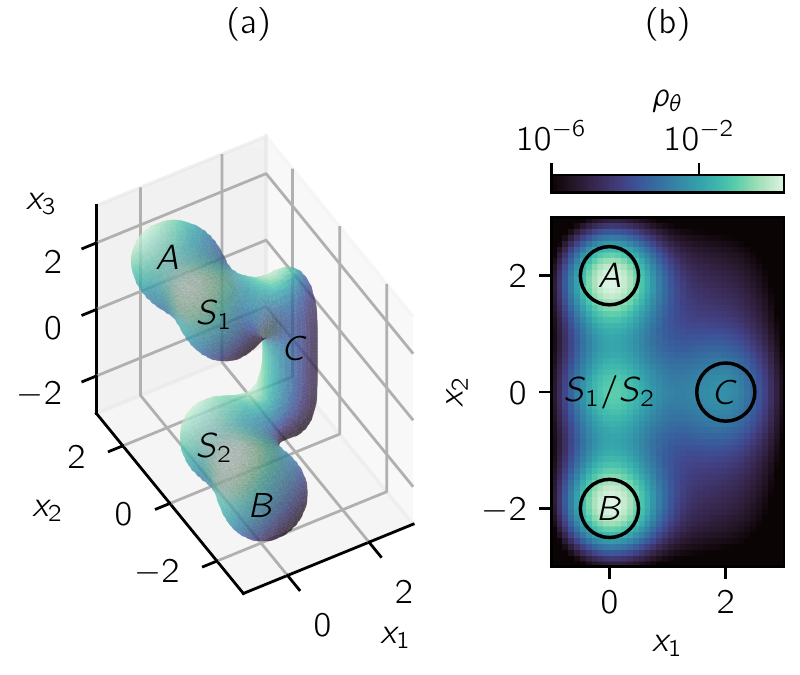}
    \caption{
        Reaction through an intermediate.
        (a) $U(x)=-3$ isosurface of \eqref{eq:example1_u}.
        (b) Marginal distribution on the CV space $(x_1,x_2)$.
    }
    \label{fig:example1_pmf}
\end{figure}

\begin{figure*}
    \includegraphics{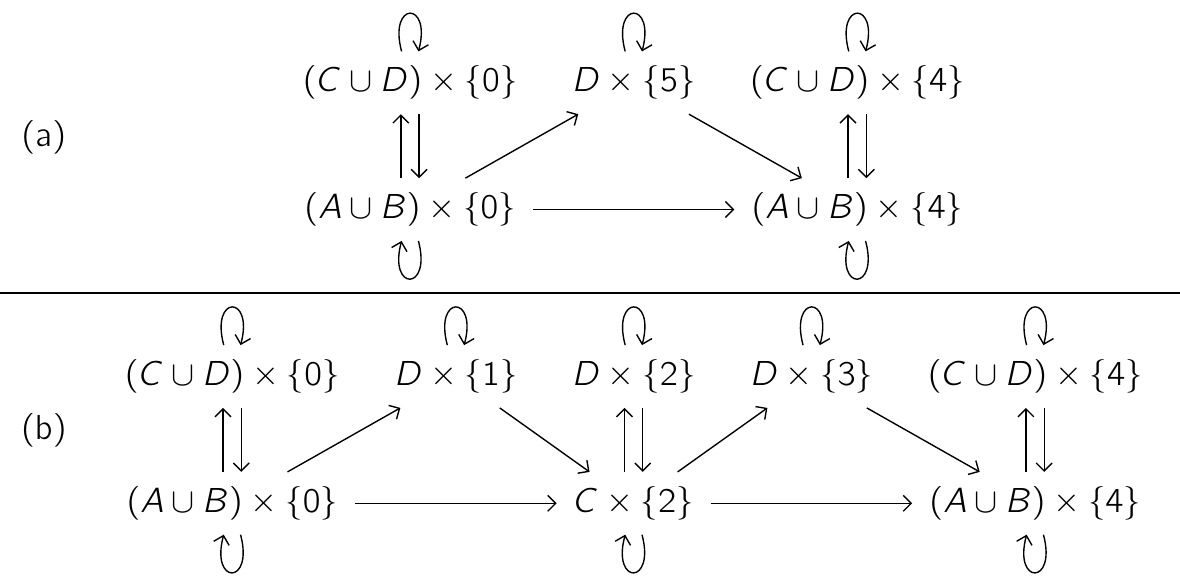}
    \caption{
        Augmented processes for reaction through an intermediate. Each arrow from one set to another indicates that $\Y{t}$ may transition to $\Y{t+\dt}$ when $\Z{t}$ is in the first set and $\Z{t+\dt}$ is in the second set.
        (a) Possible transitions of the augmented process \eqref{eq:example1_y_direct} for the uncatalyzed pathway.
        (b) Possible transitions of the augmented process \eqref{eq:example1_y} for the catalyzed pathway.
    }
    \label{fig:example1_y}
\end{figure*}

\begin{figure*}
    \includegraphics{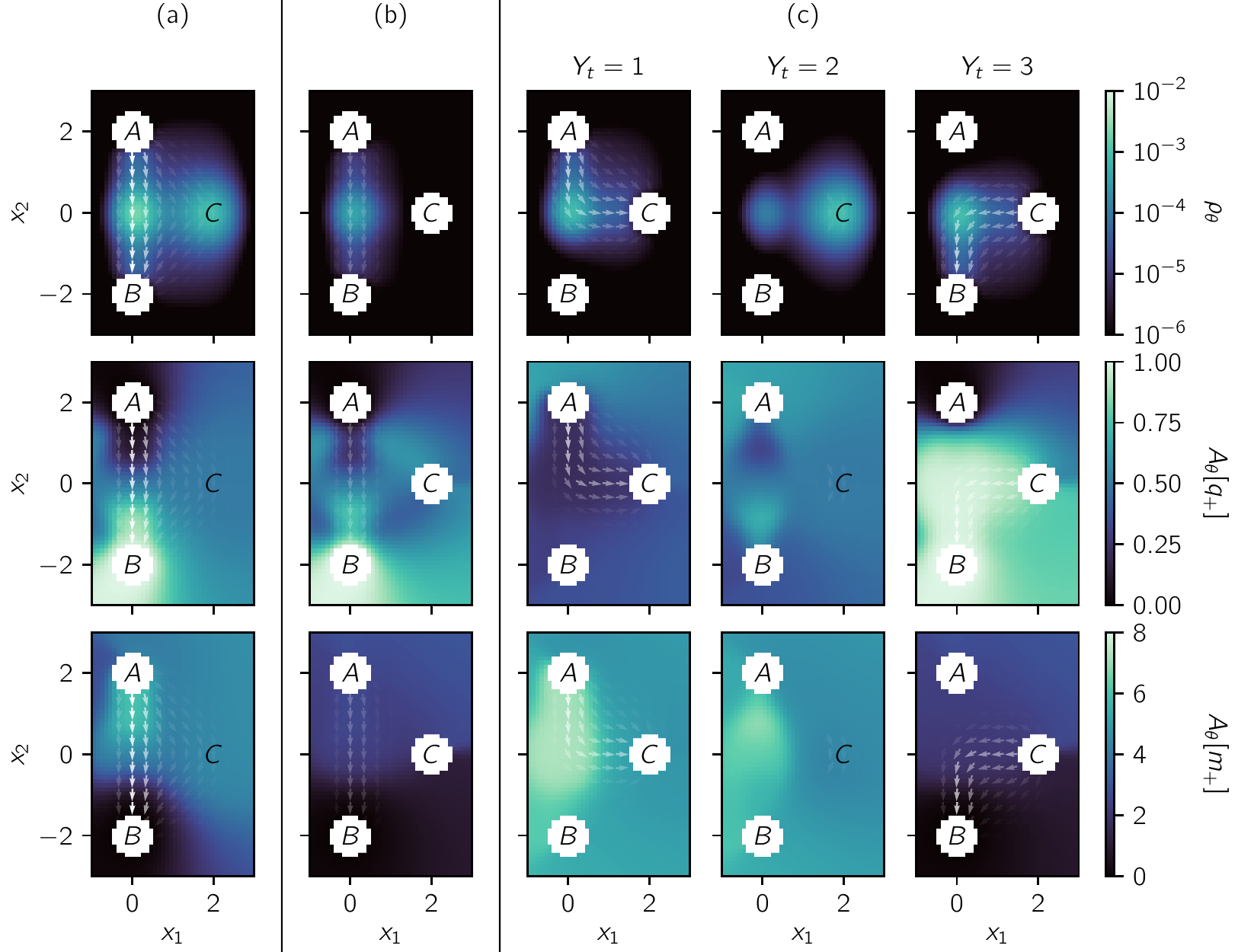}
    \caption{
        Reactive statistics for the reaction \eqref{eq:example1_tpt_w}. The reactive current (vector field) is plotted over each of the other statistics (color scale): the forward committor $\projw{\cv}[\qp]$, the reactive density $\density{\cv}$, and the conditional mean first passage time to the product $\projw{\cv}[\mtp]$. The magnitude of the reactive current is represented by the opacity of the arrows, which are comparable for the full reaction and all reaction steps. Augmented TPT statistics are computed using the augmented process \eqref{eq:example1_y}.
        (a) Reactive statistics computed using traditional TPT.
        (b) Reactive statistics for the uncatalyzed pathway computed using augmented TPT.
        (c) Reactive statistics for the catalyzed pathway computed using augmented TPT.
    }
    \label{fig:example1_stat}
\end{figure*}

In our first example, we demonstrate the use of augmented TPT to resolve individual reaction steps. We consider a reaction through an intermediate with the three-dimensional potential
\begin{equation} \label{eq:example1_u}
\begin{aligned}[b]
U(x) = 5 \Bigl[
    &
    \Bigl(\frac{x_1}{3}\Bigr)^4
    + \Bigl(\frac{x_2-1}{2}\Bigr)^4
    + \Bigl(\frac{x_3}{3}\Bigr)^4
    - e^{-x_1^2-(x_2-2)^2}
    \\
    &
    - 3e^{-(x_1+2)^2-x_2^2-(x_3-2)^2}
    - 2e^{-x_1^2-x_2^2-(x_3-2)^2}
    \\
    &
    - 3e^{-(x_1-2)^2-x_2^2-(x_3+2)^2}
    - 2e^{-x_1^2-x_2^2-(x_3+2)^2}
\Bigr],
\end{aligned}
\end{equation}
where $x = (x_1,x_2,x_3)$. We visualize the $U(x) = -3$ isosurface and the probability density on the CV space $\cv(x) = (x_1,x_2)$ in Figure~\ref{fig:example1_pmf}.

Our reaction of interest is described by the indicator function
\begin{equation} \label{eq:example1_tpt_w}
\rw(\X{r:s}) = \ind{A \times (D \cup C) \times \cdots \times (D \cup C) \times B}(\X{r:s}),
\end{equation}
where we have defined the reactant $A$, product $B$, and intermediate $C$ to be
\begin{equation}
\begin{aligned}
A & = \set{x \mid (x_1+2)^2 + x_2^2 \le 0.5^2}, \\
B & = \set{x \mid (x_1-2)^2 + x_2^2 \le 0.5^2}, \\
C & = \set{x \mid x_1^2 + (x_2-2)^2 \le 0.5^2},
\end{aligned}
\end{equation}
and $D = (A \cup B \cup C)^c$. This reaction represents, for instance, a catalyzed reaction where a substrate internal coordinate (represented by $x_1$) and the interaction of the substrate with the catalyst (represented by $x_2$) can be observed while the status of the reaction (represented by $x_3$) cannot. The observable variables form the CV space $(x_1,x_2)$, and the sets $A$, $B$, and $C$ are defined on this CV space.

There are two pathways in this reaction: uncatalyzed and catalyzed. In the uncatalyzed pathway, the system transitions from the reactant $A$ to $S_1$, then crosses directly to $S_2$ before entering the product $B$. In the catalyzed pathway, instead of directly crossing from $S_1$ to $S_2$, the system transitions from $S_1$ into the intermediate $C$ and then to $S_2$.

We select trajectories that react through each pathway by applying augmented TPT to the reaction. We define an augmented process for each of the pathways by including only the terms from \eqref{eq:k} that are involved in that pathway. For the uncatalyzed pathway, we remove all reactive trajectories that visit the intermediate $C$ by removing all terms that contain $\Y{t} \in \set{1,2,3}$ from \eqref{eq:k}, yielding
\begin{equation} \label{eq:example1_y_direct}
\xkz{t:t+\dt} =
\begin{aligned}[t]
    [ & \ind{\set{0} \times \set{0}}(\Y{t:t+\dt}) \\
    & + \ind{(D \times \set{5}) \times (D \times \set{5})}(\Z{t:t+\dt}) \\
    & + \ind{\set{4} \times \set{4}}(\Y{t:t+\dt}) \\
    & + \ind{((A \cup B) \times \set{0}) \times (D \times \set{5})}(\Z{t:t+\dt}) \\
    & + \ind{((A \cup B) \times \set{0}) \times ((A \cup B) \times \set{4})}(\Z{t:t+\dt}) \\
    & + \ind{(D \times \set{5}) \times ((A \cup B) \times \set{4})}(\Z{t:t+\dt}) ],
\end{aligned}
\end{equation}
and then select reactive trajectories using
\begin{equation}
\rw(\Z{r:s}) = \ind{(A \times \set{0}) \times (D \times \set{5}) \times \cdots \times (D \times \set{5}) \times (B \times \set{4})}(\Z{r:s}).
\end{equation}
For the catalyzed pathway, we retain only reactive trajectories that pass through the intermediate $C$ by removing all terms that contain $\Y{t} \in \set{5}$ as well as the direct transition from $(A \cup B) \times \set{0}$ to $(A \cup B) \times \set{4}$, which does not pass through $C$. This yields the augmented process
\begin{equation} \label{eq:example1_y}
\xkz{t:t+\dt} =
\begin{aligned}[t]
[ & \ind{\set{0} \times \set{0}}(\Y{t:t+\dt}) \\
& + \ind{(D \times \set{1}) \times (D \times \set{1})}(\Z{t:t+\dt}) \\
& + \ind{((C \cup D) \times \set{2}) \times ((C \cup D) \times \set{2})}(\Z{t:t+\dt}) \\
& + \ind{(D \times \set{3}) \times (D \times \set{3})}(\Z{t:t+\dt}) \\
& + \ind{\set{4} \times \set{4}}(\Y{t:t+\dt}) \\
& + \ind{((A \cup B) \times \set{0}) \times (D \times \set{1})}(\Z{t:t+\dt}) \\
& + \ind{((A \cup B) \times \set{0}) \times (C \times \set{2})}(\Z{t:t+\dt}) \\
& + \ind{(D \times \set{1}) \times (C \times \set{2})}(\Z{t:t+\dt}) \\
& + \ind{(C \times \set{2}) \times (D \times \set{3})}(\Z{t:t+\dt}) \\
& + \ind{(C \times \set{2}) \times ((A \cup B) \times \set{4})}(\Z{t:t+\dt}) \\
& + \ind{(D \times \set{3}) \times ((A \cup B) \times \set{4})}(\Z{t:t+\dt}) ].
\end{aligned}
\end{equation}
We then select reactive trajectories using
\begin{equation} \label{eq:example1_indirect}
\rw(\Z{r:s}) = \ind{(A \times \set{0}) \times D' \times \cdots \times D' \times (B \times \set{4})}(\Z{r:s}),
\end{equation}
where $D' = (D \times \set{1}) \cup ((C \cup D) \times \set{2}) \cup (D \times \set{3})$.
We show the possible transitions of \eqref{eq:example1_y_direct} in Figure~\ref{fig:example1_y}(a) and \eqref{eq:example1_y} in Figure~\ref{fig:example1_y}(b).

Our goal is to visualize the mechanism of the reaction in the CV space $(x_1,x_2)$ and quantify the relative rates of the two pathways. To this end, we examine four reactive statistics: the reactive density $\density{\cv}$, the reactive current $\current{\cv}$, the forward committor $\projw{\cv}[\qp]$, and the conditional mean first passage time to the product $\projw{\cv}[\mtp]$.

We plot the reactive statistics from traditional TPT in Figure~\ref{fig:example1_stat}(a). The reactive density $\density{\cv}$ reveals that reactive trajectories spend much of their time around $(x_1,x_2)=(0,0)$ and $(x_1,x_2)=(2,0)$, which is consistent with the presence of intermediates $S_1$/$S_2$ and $C$. The reactive current $\current{\cv}$ (vector field) suggests that the reaction is dominated by the uncatalyzed pathway, although a significant fraction does react through the catalyzed pathway. The forward committor $\projw{\cv}[\qp]$ changes rapidly around $(x_1,x_2)=(0,0)$, suggesting the presence of a bottleneck, corresponding to direct crossing from $S_1$ to $S_2$. It is almost uniform around $C$, suggesting the presence of the intermediate $C$. The conditional mean first passage time $\projw{\cv}[\mtp]$ can be interpreted in the same way as the forward committor; however, it allows us to visualize the order in which states are visited more clearly. The region below $A$ has a higher value of $\projw{\cv}[\mtp]$ than the region around $C$, which suggests that reactive trajectories usually visit the former before the latter. Together, these reactive statistics suggest a cohesive picture. The reaction is dominated by the uncatalyzed pathway, which has a bottleneck around $(x_1,x_2)=(0,0)$. Reactive trajectories may leave this pathway before the bottleneck into the catalyzed pathway, which has an intermediate $C$, and return after the bottleneck (i.e., they appear to circumvent $S_1$/$S_2$).

Some of the results from traditional TPT are misleading. For example, traditional TPT suggests that the uncatalyzed pathway is dominant, yet the total reactive flux from $A$ to $B$ is $3.0 \times 10^{-4}$, while the reactive flux for trajectories that visit $C$ is $2.2 \times 10^{-4}$, i.e., $73\%$ of trajectories go through the intermediate. This results from the restriction of the observed coordinates to $(x_1,x_2)$; in the full state space $(x_1,x_2,x_3)$, traditional TPT is capable of correctly resolving the two pathways. However, we note that even given $(x_1,x_2,x_3)$, traditional TPT cannot calculate dynamical statistics for the ensemble of trajectories that react through a particular pathway. Augmented TPT provides a solution to the overlap issue and enables the calculation of reactive statistics for individual steps of each pathway.

We first analyze the uncatalyzed pathway, which was wrongly suggested by traditional TPT to be the dominant pathway. Reactive statistics for this pathway are shown in Figure~\ref{fig:example1_stat}(b). As we would expect, the reactive density $\density{\cv}$ shows that reactive trajectories spend much of their time around $S_1$/$S_2$, and the reactive current $\current{\cv}$ suggests that reactive trajectories flow directly from $A$ to $S_1$/$S_2$ to $B$, without any notable deviation to the vicinity of $C$. On the uncatalyzed pathway, the forward committor $\projw{\cv}[\qp]$ changes rapidly around $(x_1,x_2)=(0,0)$ due to the transition from $S_1$ to $S_2$. Off the uncatalyzed pathway, the forward committor has a higher value closer to $A$ and a lower value closer to $B$. This is surprising and results from slight differences in the reactive density at different values of $x_3$. The conditional mean first passage time $\projw{\cv}[\mtp]$ rapidly decreases near $S_1$/$S_2$, suggesting the same single bottleneck. 

We now analyze the catalyzed pathway through the intermediate $C$, which dominates the rate. We select trajectories that react through this pathway using \eqref{eq:example1_indirect}. Augmented TPT enables us to split the pathway into individual steps, and so resolve the structure of each reaction step. The first step ($\Y{t}=1$) starts when the reactive trajectory leaves the reactant $A$ and ends when it first enters the intermediate $C$. The second step ($\Y{t}=2$) starts at the first time the reactive trajectory enters $C$, and ends at the last time the reactive trajectory leaves $C$. The third step ($\Y{t}=3$) starts when the reactive trajectory last leaves the intermediate $C$ and ends when it enters the product $B$. As we now explain, separating the catalyzed pathway into these steps leads to reactive statistics that lead to a different interpretation than those from traditional TPT.

In the first step, the reactive density $\density{\cv}$ and reactive current $\current{\cv}$ clearly show that most reactive trajectories flow through an intermediate near $(x_1,x_2)=(0,0)$, in this case $S_1$, rather than a more direct path from $A$ to $C$, as suggested by the reactive current from traditional TPT. Likewise, in the last step, they show that most reactive trajectories flow through an intermediate near $(x_1,x_2)=(0,0)$, in this case $S_2$. The absence of any significant reactive current in the second step, along with the high reactive density near $C$, suggests that reactive trajectories predominantly remain in $C$ during this step, with a few trajectories transitioning back and forth to $S_1$/$S_2$, where there is a lower value of reactive density. The reactive current from traditional TPT is misleading because the flows from $S_1$ to $C$ and from $C$ to $S_2$ cancel each other, since $S_1$ and $S_2$ overlap in the CV space.

The forward committor $\projw{\cv}[\qp]$ on the catalyzed pathway is uniformly low in the first step and uniformly high in the third step. This suggests that the main bottleneck occurs in the second step, where $\projw{\cv}[\qp] \approx 0.5$ around $C$. The abrupt changes between steps suggest that the dynamics of the variables not captured within the CV space are influential in determining whether the reaction occurs. For the second step, we note that the low value of $\projw{\cv}[\qp]$ below $A$ and high value above $B$ reflect the full three-dimensional potential (Figure~\ref{fig:example1_pmf}(a)). In traditional TPT, the high value of $\projw{\cv}[\qp]$ in the first step and the low value in the third step cancel, giving rise to the apparent rapid change near $(x_1,x_2)=(0,0)$.

In the first step, $\projw{\cv}[\mtp]$ decreases from $(x_1,x_2)=(0,0)$ to $C$, suggesting the presence of a bottleneck between intermediates $S_1$ to $C$. The same holds for the second step, suggesting that if the system crosses back to an intermediate near $(x_1,x_2)=(0,0)$, it needs to overcome the same bottleneck to return to $C$. In the third step, $\projw{\cv}[\mtp]$ decreases from $(x_1,x_2)=(0,0)$ to $B$, marking the transition from $S_2$ to $B$. We note that this decrease occurs at a slightly lower value than in the uncatalyzed pathway. This separation of the two bottlenecks lies in contrast with $\projw{\cv}[\mtp]$ from traditional TPT, where the superposition of the two bottlenecks creates an apparent bottleneck near $(x_1,x_2)=(0,0)$, which conflates the dynamics of the catalyzed and uncatalyzed pathways.

Overall, we see that the statistics from traditional TPT qualitatively resemble a superposition of those for the uncatalyzed pathway and those associated with the second step of the catalyzed pathway, in which the system is mainly localized at $C$. The important contributions from the first and third steps of the catalyzed pathway mask each other in traditional TPT.

\subsection{Reaction with Multiple Pathways}

\begin{figure}
\includegraphics{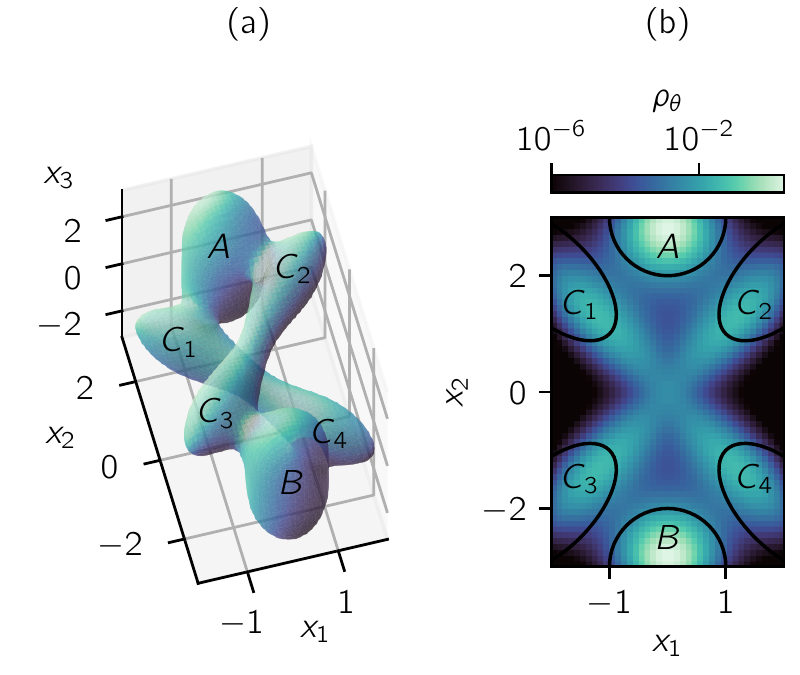}
\caption{
Reaction with multiple pathways.
(a) $U(x)=-3$ isosurface of \eqref{eq:example2_u}.
(b) Marginal distribution on the CV space $(x_1,x_2)$.
}
\label{fig:example2_u}
\end{figure}

\begin{figure*}
    \includegraphics{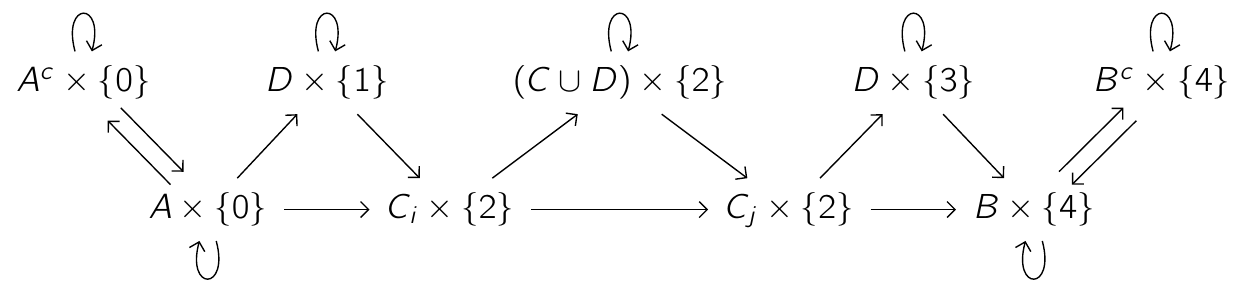}
    \caption{Possible transitions for the augmented process of the reaction with multiple pathways, defined by \eqref{eq:example2_y}.}
    \label{fig:example2_k}
\end{figure*}

\begin{figure*}
\includegraphics{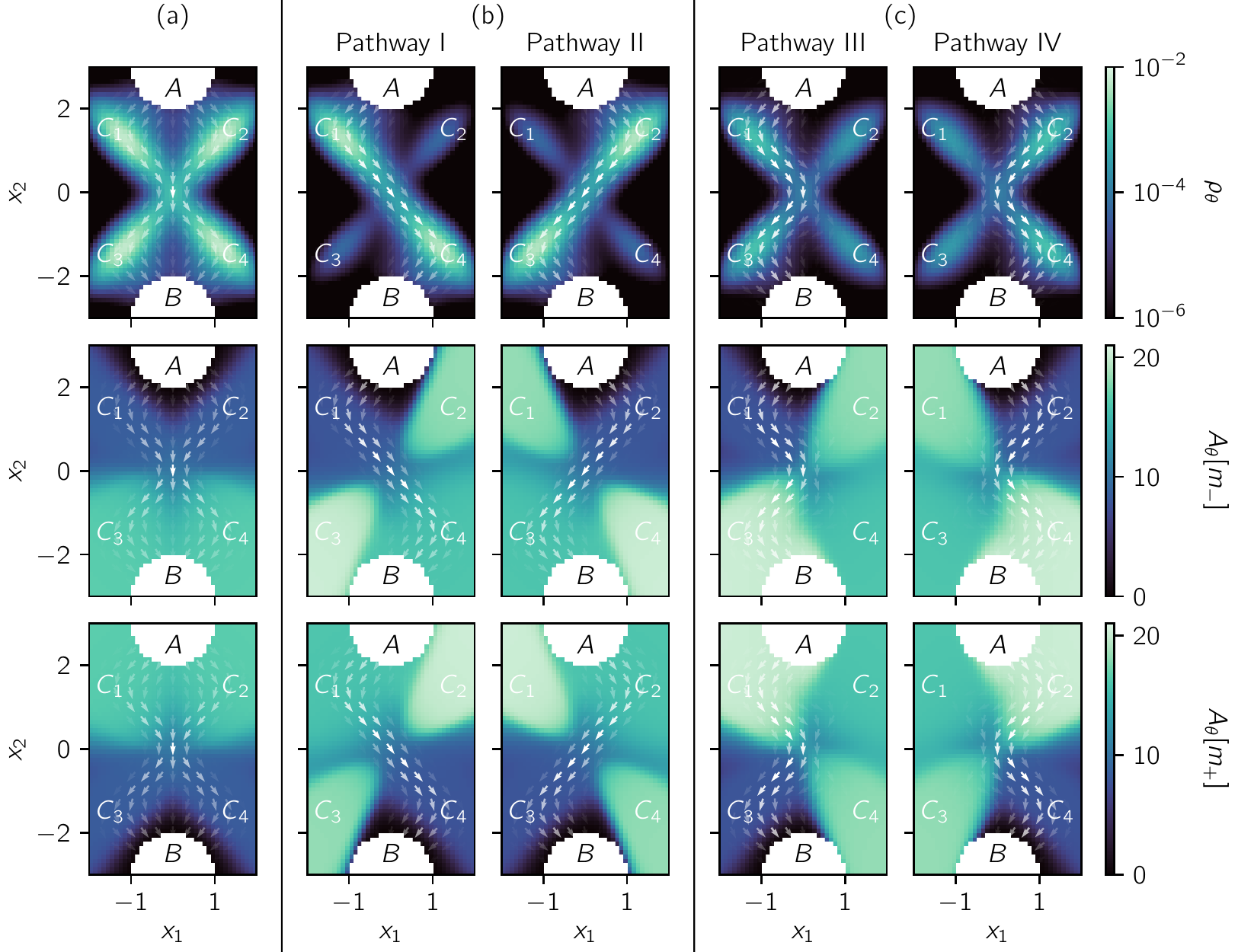}
\caption{
    Reactive statistics for the reaction with multiple pathways. The reactive current (vector field) is plotted over each of the other statistics. Note that the magnitude of the reactive current is represented by the opacity of the arrows, and that they are normalized differently for each reaction and pathway. The other reactive statistics (color scale) are the reactive density $\density{\cv}$, the conditional mean last passage time from the reactant $\projw{\cv}[\mtm]$, and the conditional mean first passage time to the product $\projw{\cv}[\mtp]$.
    (a) Reactive statistics from traditional TPT.
    (b) Reactive statistics from augmented TPT for major pathways I and II.
    (c) Reactive statistics from augmented TPT for minor pathways III and IV.
}
\label{fig:example2_stat}
\end{figure*}

For our second example, we demonstrate the use of augmented TPT to separate pathways which overlap in the CV space. We consider overdamped Langevin dynamics on the three-dimensional potential
\begin{equation} \label{eq:example2_u}
\begin{aligned}[b]
U(x) =
    5 \Bigl[
    &
    \Bigl(\frac{x_1}{3}\Bigr)^4
    + \Bigl(\frac{x_2}{3}\Bigr)^4
    + \Bigl(\frac{x_3}{2}\Bigr)^4
    + e^{-x_1^2-x_2^2-x_3^2}
    \\
    &
    - 3e^{-(x_1-2)^2-(x_2-2)^2}
    - 2e^{-x_2^2+(x_3-1)^2}
    \\
    &
    - 3e^{-(x_1+2)^2-(x_2+2)^2}
    - 2e^{-x_1^2-(x_3+1)^2}
    \Bigr],
\end{aligned}
\end{equation}
where $x = (x_1,x_2,x_3)$. As previously, these dynamics satisfy the Fokker--Planck equation in \eqref{eq:overdamped}. The $U(x) = -3$ isosurface for this potential is shown in Figure~\ref{fig:example2_u}(a), and the probability distribution on the $(x_1,x_2)$ coordinates, which we use as CVs, is shown in Figure~\ref{fig:example2_u}(b). We define the reactant $A$, product $B$, and intermediates $C_1$, $C_2$, $C_3$, and $C_4$ to be
\begin{equation}
\begin{aligned}
A & = \set{x \mid x_1^2 + (x_2-3)^2 \le 1}, \\
B & = \set{x \mid x_1^2 + (x_2+3)^2 \le 1}, \\
C_1 & = \set{x \mid (x_1+x_2)^2 + (x_1-x_2+4)^2/4 \le 1}, \\
C_2 & = \set{x \mid (x_1+x_2-4)^2/4 + (x_1-x_2)^2 \le 1}, \\
C_3 & = \set{x \mid (x_1+x_2+4)^2/4 + (x_1-x_2)^2 \le 1}, \\
C_4 & = \set{x \mid (x_1+x_2)^2 + (x_1-x_2-4)^2/4 \le 1}.
\end{aligned}
\end{equation}
We also define the sets $C = C_1 \cup C_2 \cup C_3 \cup C_4$ and $D = (A \cup B \cup C)^c$. The reaction of interest is specified through the indicator function
\begin{equation} \label{eq:example2_w}
    \rw(\X{r:s}) = \ind{A \times (D \cup C) \times \cdots \times (D \cup C) \times B}(\X{r:s}).
\end{equation}

We use the intermediates to define pathways. This is advantageous because it is not possible to divide the space into regions corresponding to different pathways owing to overlap. We define four pathways: major pathways I and II, and minor pathways III and IV. We define pathway I to first hit intermediate $C_i = C_1$ after leaving $A$ and last hit intermediate $C_j = C_4$ before hitting $B$. We likewise define pathway II with $(C_i,C_j) = (C_2,C_3)$, pathway III with $(C_i,C_j) = (C_1,C_3)$, and pathway IV with $(C_i,C_j) = (C_2,C_4)$.

To define means of selecting these pathways, we label a trajectory before the reaction with $\Y{t} = 0$ and after the reaction with $\Y{t} = 4$. When the reaction is in process, we label times before the reactive trajectory first enters $C_i$ with $\Y{t} = 1$, after the reactive trajectory last exits $C_j$ with $\Y{t} = 3$, and between those times with $\Y{t} = 2$. Then, we select pathways using
\begin{equation}
    \rw(\Z{r:s}) = \ind{(A \times \set{0}) \times D' \cdots D' \times (B \times \set{4})}(\Z{r:s})
\end{equation}
where $D' = (D \times \set{1}) \cup ((D \cup C) \times \set{2}) \cup (D \times \set{3})$. This choice of $\Y{t}$ corresponds to
\begin{equation} \label{eq:example2_y}
\xkz{t:t+\dt} =
\begin{aligned}[t]
[
    & \ind{\set{0} \times \set{0}}(\Y{t:t+\dt}) \\
    & + \ind{(D \times \set{1}) \times (D \times \set{1})}(\Z{t:t+\dt}) \\
    & + \ind{((C \cup D) \times \set{2}) \times ((C \cup D) \times \set{2})}(\Z{t:t+\dt}) \\
    & + \ind{(D \times \set{3}) \times (D \times \set{3})}(\Z{t:t+\dt}) \\
    & + \ind{\set{4} \times \set{4}}(\Y{t:t+\dt}) \\
    & + \ind{(A \times \set{0}) \times (D \times \set{1})}(\Z{t:t+\dt}) \\
     & + \ind{(A \times \set{0}) \times (C_i \times \set{2})}(\Z{t:t+\dt}) \\
     & + \ind{(D \times \set{1}) \times (C_i \times \set{2})}(\Z{t:t+\dt}) \\
    & + \ind{(C_j \times \set{2}) \times (D \times \set{3})}(\Z{t:t+\dt}) \\
    & + \ind{(C_j \times \set{2}) \times (B \times \set{4})}(\Z{t:t+\dt}) \\
    & + \ind{(D \times \set{3}) \times (B \times \set{4})}(\Z{t:t+\dt})
].
\end{aligned}
\end{equation}
The first five terms of \eqref{eq:example2_y} denote the sets in which $\X{t}$ may be for each of the labels $\set{0,1,2,3,4}$, and the remaining six terms describe the permitted transitions between the labels. We represent \eqref{eq:example2_y} visually in Figure~\ref{fig:example2_k}. For instance, the seventh term, $\ind{(A \times \set{0}) \times (C_i \times \set{2})}(\Z{t:t+\dt})$, corresponds to the single timestep transition from $A$ to $C_i$ and associates this with a change in the label from $\Y{t} = 0$ to $\Y{t} = 2$.

We determine the reactive flux associated with each pathway and compare it with the total reactive flux. For the reaction specified by \eqref{eq:example2_w}, the total reactive flux is $7.5 \times 10^{-4}$. Each of the major pathways has a reactive flux of $2.9 \times 10^{-4}$, which is $39\%$ of the total reactive flux. Each of the minor pathways has a reactive flux of $6.4 \times 10^{-4}$, which is $8.5\%$ of the total reactive flux. These four pathways thus give rise to $95\%$ of the total reactive flux, and so are representative of the majority of reactive trajectories.  The remaining 5\% results from trajectories that do not conform to these pathways (e.g., ones that pass through only a single intermediate).

In Figure~\ref{fig:example2_stat}, we plot four reactive statistics: the reactive density $\density{\cv}$, the reactive current $\current{\cv}$, the conditional mean last passage time from the reactant $\projw{\cv}[\mtm]$, and the conditional mean first passage time to the product $\projw{\cv}[\mtp]$.

Reactive statistics from traditional TPT are shown in Figure~\ref{fig:example2_stat}(a). From the reactive density $\density{\cv}$, we observe that reactive trajectories spend most of their time in the X-shaped region that connects the intermediates. The reactive current $\current{\cv}$ suggests that the majority of the reactive trajectories flow from the reactant $A$ to either $C_1$ or $C_2$, then to either $C_3$ or $C_4$ via $(x_1,x_2)=(0,0)$, and lastly to the product $B$. Importantly, even if we consider the full state space $(x_1,x_2,x_3)$, we cannot determine the relative weights of these four possible pathways, because the pathways are composed of segments that belong to multiple pathways (e.g., the first half of pathway III overlaps with pathway I and the second half of pathway III overlaps with pathway II) and the transitions through  $(x_1,x_2,x_3)=(0,0,0)$ along pathways III and IV occur in opposite directions. Other quantities calculated using traditional TPT have the same issue. Both $\projw{\cv}[\mtm]$ and $\projw{\cv}[\mtp]$ are unable to distinguish between the pathways and only indicate the presence of a bottleneck between $C_1 \cup C_2$ and $C_3 \cup C_4$.

In Figure~\ref{fig:example2_stat}(b), we visualize the reactive statistics for the major pathways. As pathway I and pathway II are mirror images of one another, we discuss only pathway I. The reactive current $\current{\cv}$ clearly shows that the system transitions directly between on-pathway intermediates, from $A$ to $C_1$ to $C_4$ to $B$. The reactive density $\density{\cv}$ corroborates this picture, with relatively little density in $C_2$ and $C_3$ compared to $C_1$ and $C_4$. We observe that $\projw{\cv}[\mtm]$ and $\projw{\cv}[\mtp]$ are highest near $C_2$ and $C_3$, suggesting that these configurations are dynamically disconnected from the main flow of the reactive trajectories. The transition from $C_1$ to $C_4$ is accompanied by an abrupt increase in $\projw{\cv}[\mtm](\cvpt)$ and an abrupt decrease in $\projw{\cv}[\mtp]$, which shows that a transition bottleneck is traversed. We note that the sharp increase in $\projw{\cv}[\mtm]$ from $A$ to $C_1$ and the sharp decrease in $\projw{\cv}[\mtp]$ from $C_4$ to $B$ imply the presence of bottlenecks between each of these pairs of states.

The minor pathways in Figure~\ref{fig:example2_stat}(c) result from trajectories that switch between the major pathways. Since the two minor pathways are mirror images of each other, we discuss only pathway III, which involves a switch from pathway I to pathway II. In contrast with the major pathways, the reactive density $\density{\cv}$ indicates that the system is likely to visit off-pathway intermediates $C_2$ and $C_4$ on its way from $C_1$ to $C_3$. The conditional mean last passage time $\projw{\cv}[\mtm]$ from the reactant is nearly identical to that of pathway I, with higher values near $C_2$ and $C_3$ and lower values near $C_1$ and $C_4$. However, the conditional mean first passage time $\projw{\cv}[\mtp]$ to the product is nearly identical to that of pathway II. In conjunction with $\density{\cv}$, the slight increase in $\projw{\cv}[\mtm]$ from $C_1$ to $C_4$ suggests that these intermediates readily interconvert, and similarly for the slight increase in $\projw{\cv}[\mtp]$ from $C_3$ to $C_2$. The larger change in $\projw{\cv}[\mtm]$ and $\projw{\cv}[\mtp]$ between $C_2 \cup C_3$ and $C_1 \cup C_4$ implies a bottleneck between $C_2 \cup C_3$ and $C_1 \cup C_4$. This bottleneck is significant because the transition from $C_1$ to $C_3$ must occur for this pathway. As with the major pathways, there are also bottlenecks between $A$ and $C_1$, and $C_3$ and $B$.

\section{Discussion} \label{sec:discussion}

In this paper, we introduced an augmented process that labels sequences of events. This process enabled us to write statistics that depend on knowledge of past and future events in terms of quantities that are local in time and in turn to extend the TPT framework. We demonstrated how this framework can be used to separate statistics of competing pathways in reactions with intermediates to reveal features of mechanisms that are not apparent from traditional TPT analyses. Our framework can also be used to treat new classes of reactions that are not amenable to TPT analyses. For instance, reactions with the same reactant and product states, such as cycles of oscillators and excitable systems, can be handled using augmented TPT but not traditional TPT.

Our framework generalizes a previous extension of TPT \cite{miron2021} and history augmented approaches for computing rates \cite{suarez2016,vanden2009exact,dickson2009separating} and reactive statistics \cite{vani2022}. The augmented process that we introduce is distinct from that in ref.~\onlinecite{helfmann2020}, in which the state space is expanded to include a time variable to treat time-dependent processes, including transient relaxations and systems with periodically varying dynamics. As a result, the two approaches can be combined to treat sequences of events of finite-time processes.

Our focus here was on establishing the conceptual framework for augmented TPT, and the examples that we showed were sufficiently simple that the Fokker--Planck equations defining their dynamics could be numerically integrated in the variables by quadrature. The dynamics of models with larger numbers of variables must instead be sampled through simulations that generate stochastic realizations of trajectories (i.e., the dynamics of the variables are numerically integrated in time). Because, like traditional TPT, the framework casts statistics in terms of quantities that are local in time, we can extend methods that compute reactive statistics from short trajectories \cite{thiede2019,strahan2021,vani2022}. Such efforts are underway.

\section*{Acknowledgments}

We acknowledge Adam Antoszewski, Spencer Guo, John Strahan, and Bodhi Vani for useful discussions. We thank the Research Computing Center at the University of Chicago for computational resources. This work was supported by National Institutes of Health award R35 GM136381 and National Science Foundation award DMS-2054306.

\appendix

\section{Finite Difference Scheme} \label{appendix:fdm}

For a time-reversible drift-diffusion process
\begin{equation} \label{eq:ito-process}
\frac{\partial \Px[\X{t}]}{\partial t} = \nabla \cdot (\Px[\X{t}] \nabla U(\X{t})) + \nabla^2 \Px[\X{t}]
\end{equation}
with stationary distribution $\pi(\X{t}) \propto \exp(-U(\X{t}))$, the infinitesimal generator can be used to compute expectations forward-in-time as
\begin{equation}
\frac{\partial \Ex[f(\X{t+\dt}) \mid \X{t}]}{\partial \dt}
= -\nabla U(\X{t}) \cdot \nabla f(\X{t}) + \nabla^2 f(\X{t}).
\end{equation}

To evaluate expectations by quadrature, we adapt the finite difference scheme from ref.~\onlinecite{thiede2019}, which we reproduce here. We approximate \eqref{eq:ito-process} as a discrete time Markov jump process with time step $\dt$ on a grid with uniform spacing $\epsilon$. For a small change $\epsilon_i$ in the direction of the $i$\textsuperscript{th} coordinate with magnitude $\epsilon$, we substitute $-\nabla U(\X{t}) = \nabla \pi(\X{t}) / \pi(\X{t})$ and then make the approximation
\begin{equation}
\begin{aligned}
\MoveEqLeft[1] \frac{\E[f(\X{t+\dt}) \mid \X{t}] - f(\X{t})}{\dt} \\
& \approx
\begin{aligned}[t]
& \frac{1}{2} \sum_i \frac{(\pi(\X{t}+\epsilon_i)-\pi(\X{t}))/\epsilon}{(\pi(\X{t}+\epsilon_i)+\pi(\X{t}))/2} \left[\frac{f(\X{t}+\epsilon_i)-f(\X{t})}{\epsilon}\right] \\
& + \frac{1}{2} \sum_i \frac{(\pi(\X{t})-\pi(\X{t}-\epsilon_i))/\epsilon}{(\pi(\X{t})+\pi(\X{t}-\epsilon_i))/2} \left[\frac{f(\X{t})-f(\X{t}-\epsilon_i)}{\epsilon}\right] \\
& + \sum_i \frac{f(\X{t}+\epsilon_i)+f(\X{t}-\epsilon_i)-2f(\X{t})}{\epsilon^2}.
\end{aligned}
\end{aligned}
\label{eq:L_approx}
\end{equation}
Alternatively, we can write
\begin{align}
\MoveEqLeft[1] \frac{\Ex[f(\X{t+\dt}) \mid \X{t}] - f(\X{t})}{\dt}
\nonumber \\
& = \frac{1}{\dt} \left(\frac{\int \Px[\X{t:t+\dt}] f(\X{t+\dt}) \odif{\X{t+\dt}}}{\Px[\X{t}]} - f(\X{t})\right) \\
& \approx \begin{aligned}[t]
    \frac{1}{\dt} \Bigl[
        \Bigl( & P(\X{t},\X{t}) f(\X{t}) \\
            & + \sum_i P(\X{t},\X{t}+\epsilon_i) f(\X{t}+\epsilon_i) \\
            & + \sum_i P(\X{t},\X{t}-\epsilon_i) f(\X{t}-\epsilon_i)
        \Bigr) / \pi(\X{t})
    - f(\X{t}) \Bigr],
\end{aligned}
\label{eq:P_approx}
\end{align}
where $P(\X{t:t+\dt})$ represents the approximation of $\Px[\X{t:t+\dt}]$ on the grid. Above, the first equality follows from the definition of conditional expectation, and the second assumes that all transitions within time $\Delta$ are to neighboring grid points.

By matching terms between \eqref{eq:P_approx} and \eqref{eq:L_approx}, the only nonzero entries of $P(x,x')$ are
\begin{align}
P(x,x+\epsilon_i) & = \frac{2 \dt}{\epsilon^2} \frac{1}{1/\pi(x) + 1/\pi(x+\epsilon_i)}, \\
P(x,x-\epsilon_i) & = \frac{2 \dt}{\epsilon^2} \frac{1}{1/\pi(x) + 1/\pi(x-\epsilon_i)}, \\
P(x,x) & = \pi(x) - \sum_i [P(x,x+\epsilon_i) + P(x,x-\epsilon_i)].
\end{align}
Then, we can use the above expressions to estimate expectations using
\begin{align}
\Ex[f(\X{t-\dt:t}) \mid \X{t}] & \approx \frac{\sum_{\X{t-\dt}} P(\X{t-\dt:t}) f(\X{t-\dt:t})}{\pi(\X{t})},
\label{eq:fdm_m} \\
\Ex[f(\X{t:t+\dt}) \mid \X{t}] & \approx \frac{\sum_{\X{t+\dt}} P(\X{t:t+\dt}) f(\X{t:t+\dt})}{\pi(\X{t})},
\label{eq:fdm_p}\\
\Ex[f(\X{t:t+\dt})] & \approx \sum_{\X{t:t+\dt}} P(\X{t:t+\dt}) f(\X{t:t+\dt}),
\label{eq:fdm_exp} 
\end{align}
where the sums are over points on the grid.

\section*{References}

%

\end{document}